\documentclass[journal, twocolumn]{IEEEtran}
\ifCLASSINFOpdf
  \usepackage[pdftex]{graphicx}
  \graphicspath{{../pdf/}{../jpeg/}}
  \DeclareGraphicsExtensions{.pdf,.jpeg,.png}
\else
  \usepackage[dvips]{graphicx}
  \graphicspath{{../eps/}}
  \DeclareGraphicsExtensions{.eps}
\fi
\usepackage{color}
\usepackage{epsfig}
\usepackage{booktabs}
\usepackage{epstopdf}
\usepackage{amssymb}
\usepackage{mathtools}
\usepackage{algorithmic}
\usepackage{array}
\usepackage{mdwmath}
\usepackage{mdwtab}
\usepackage{bm}
\usepackage{mathrsfs}
\usepackage{epsfig}
\usepackage{subfigure}
\usepackage{algorithmic}
\usepackage{amsmath}
\usepackage{mathrsfs}
\usepackage{cite}
\usepackage{url}

\DeclareGraphicsExtensions{pst,eps}
\begin{document}

\title{Meta-Reinforcement Learning Based Resource Allocation for Dynamic V2X Communications}
\author{Yi~Yuan,~Gan~Zheng,~\IEEEmembership{Fellow,~IEEE,}  Kai-Kit Wong,~\IEEEmembership{Fellow,~IEEE,}
and Khaled B. Letaief, ~\IEEEmembership{Fellow,~IEEE}
\thanks{Y. Yuan was with the Wolfson School of Mechanical, Electrical and Manufacturing Engineering, Loughborough University, Loughborough, LE11 3TU, UK, and now is with the 5G Innovation Center, Institute of Communication Systems, University of Surrey, Guildford GU2 7XH, U.K (E-mail: yy0007@surrey.ac.uk).}
\thanks{G. Zheng is with the Wolfson School of Mechanical, Electrical and Manufacturing Engineering, Loughborough University, Loughborough, LE11 3TU, UK (E-mail: g.zheng@lboro.ac.uk).}
\thanks{K.-K. Wong is with the Department of Electronic and Electrical Engineering, University College London, London, WC1E 6BT, UK (Email: kai-kit.wong@ucl.ac.uk).}
\thanks{K. B. Letaief is with the Department of Electronic and Computer Engineering, Hong Kong University of Science and Technology,
Clear Water Bay, Hong Kong, and is also with Peng Cheng Laboratory, Shenzhen 518066, China (E-mail: eekhaled@ust.hk).} }
\maketitle

\begin{abstract}
This paper studies the allocation of shared resources between vehicle-to-infrastructure (V2I) and vehicle-to-vehicle (V2V) links in vehicle-to-everything (V2X) communications. In existing algorithms, dynamic vehicular environments and quantization of continuous power become the bottlenecks for providing an effective and timely resource allocation policy. In this paper, we develop two algorithms to deal with these difficulties. First, we propose a  deep reinforcement learning (DRL)-based resource allocation  algorithm to improve the performance of both V2I and V2V links. Specifically, the algorithm uses deep Q-network (DQN) to solve the sub-band assignment and   deep deterministic policy-gradient (DDPG) to solve the continuous power allocation problem. Second, we propose a meta-based DRL algorithm to enhance the fast adaptability of the resource allocation policy in the dynamic environment. Numerical results   demonstrate that the proposed DRL-based algorithm can significantly improve the performance compared to the DQN-based algorithm that quantizes continuous power. In addition, the proposed meta-based DRL algorithm can achieve the required fast adaptation in the new environment with limited experiences.
\end{abstract}

\begin{IEEEkeywords}
Vehicular communications, meta-learning, deep reinforcement learning, DDPG.
\end{IEEEkeywords}
 
\section{Introduction}
Vehicle-to-everything (V2X) communications has been recognized as a key technology to support safe and efficient intelligent transportation services \cite{wang2018networking}. The cellular V2X technique has been widely developed and deployed by 5G automotive association (5GAA) due to its ability on providing better coverage and quality of service (QoS).  This technique includes two important communications modes: vehicle-to-infrastructure (V2I) and   vehicle-to-vehicle (V2V). V2I communications focus on the high data rate service and V2V communications focus on the safety-critical messages delivery \cite{lu2014connected}. In order to satisfy the stringent requirements in V2X communications, the cellular V2X technique is required to provide simultaneous V2I and V2V communications using the shared resource pool. Therefore, how to manage the interference and support the coexistence of V2I and V2V connections within a limited frequency spectrum becomes an important problem in V2X communications.

The traditional optimization approaches have been used to design the model-oriented algorithms to cope with the resource allocation and interference management in V2X communications \cite{ren2015powe,sun2015radio,sun2015cluster,liang2017resource,liang2018graph,mei2018latency,liu2018ultra}. In \cite{ren2015powe}, a resource allocation strategy was proposed to improve the throughput based on the formulated device-to-device (D2D)-based vehicle communications framework. In \cite{sun2015radio}, Sun et al. proposed a radio resource management (RRM) algorithm to guarantee the latency and reliability requirements of the D2D-based V2X system. Based on the similar V2X framework, a three-stage RRM algorithm was proposed to allocate the spectrum, which is shared by the different V2I links and V2V pairs \cite{sun2015cluster}. An optimization problem was formulated to design an efficient spectrum and power allocation algorithm for a D2D-enabled vehicular system, which only considers the slowly varying large-scale fading channel \cite{liang2017resource}. Furthermore, a graph partition algorithm was exploited to control the interference caused by V2V links \cite{liang2018graph}. Besides, the impacts of the queueing latency on the throughput and reliability were investigated in \cite{mei2018latency} and \cite{liu2018ultra}.

Although traditional optimization methods have been widely used to solve wireless communication problems, there exist some limitations of such methods on solving resource allocations in V2X communications. First, due to the high mobility feature of vehicles, it is hard to obtain the precise channel state information (CSI) of rapidly varying mobile links with low signaling overhead. Second, the iterative algorithms designed by the traditional optimization methods cannot make the fast decision on the resource allocation in a rapidly varying channel scenario since such algorithms require iterative calculations for each channel response. Third, some requirements for V2X communications are hard to solve in a mathematically exact manner, such as the reliability requirement of V2V links to deliver the packets within a time limit.

 Machine learning has been considered as a promising technique to tackle   challenges in traditional optimization methods \cite{zhang2019deepp,zappone2019wireless}. Supervised learning and reinforcement learning (RL) are two widely used machine learning techniques in wireless communications. Supervised learning heavily relies on labelled data \cite{mohri2018foundations}, i.e., the optimal solutions, which are not easy to obtain. Different from supervised learning, RL aims to learn a better policy on decision-making through exploration and exploitation \cite{sutton2018reinforcement}. By combining RL with DNN, deep RL (DRL) is designed in \cite{li2017deep}, which extends RL to deal with the large state-action space. DRL has demonstrated its strong ability in solving a more complex Markov decision process (MDP) with a high-dimensional state-action space in wireless communications applications \cite{luong2019applications}, especially for decision-making problems in V2X communications \cite{zhang2019mobile, liang2019deep}. DRL has been widely adopted to address the more complex resource management problems in V2X communications scenarios \cite{he2017integrated,atallah2018scheduling,zhang2018artificial,ye2019deep,liang2019spectrum,zhang2019deep}. In order to improve the performance of the next generation vehicular networks, the authors proposed a deep Q-network (DQN)-based algorithm to control the complex resources via an integrated framework \cite{he2017integrated}. The DRL technique was applied to address the resource allocation in the Internet of Vehicle (IoV) case, which adopts the concept of Internet of Things (IoT) to vehicle communications scenarios to improve the road safety and satisfy the ubiquitous connectivity requirements \cite{atallah2018scheduling,zhang2018artificial}. In \cite{atallah2018scheduling}, the safety and QoS issues in the IoV case with the battery-powered vehicles were solved by exploiting DQN to learn an optimal mapping from the current characteristics of the underlying model to the scheduling policy. A DQN-based algorithm was designed to overcome the dynamic topology and time-varying spectrum states by learning the optimal scheduling policy in a cognitive radio-based vehicular network \cite{zhang2018artificial}. The resource allocation problem, which considers the latency of V2V links and the sum rate of V2I links, was investigated in \cite{ye2019deep} and \cite{liang2019spectrum} by using the single-agent and multi-agent approaches, respectively. A centralized learning and distributed implementation are adopted to select the resources based on the proposed multi-agent RL algorithm \cite{liang2019spectrum}. The decentralized DQN-based algorithm was proposed to find the optimal sub-band and power level selection in the unicast and broadcast V2X scenario \cite{ye2019deep}. In order to enhance the reliability of safety-critical messages delivery in V2V links, a two-timescale federated DRL-based semi-decentralized algorithm was proposed to optimize the selection of transmission mode and resources in V2X communications \cite{zhang2019deep}.

Although DRL has shown its advantages of performance and complexity compared to the traditional optimization approaches, there exist two main drawbacks for the existing DRL-based algorithms proposed in \cite{he2017integrated,atallah2018scheduling,zhang2018artificial,ye2019deep,liang2019spectrum,zhang2019deep} on addressing the decision-making problems in the real-time V2X communications. First, DQN is used as the main technique to handle the discrete-continuous hybrid action space by discretizing continuous actions, which causes quantization error and degrades the performance since the output of DQN relies on the selection of the best action. In addition, the high dimensional quantization on the continuous action space will cause an exponential increase of the computational complexity. Second, the current DRL-based algorithms are designed based on the assumption that there is no change between the training and testing environments. However, such an assumption is impractical in V2X communications scenarios due to the high mobility and dynamic features of the vehicular environment. As a result, the existing DRL-based algorithm may cause the mismatch issue when the environment changes, which means such algorithms cannot make the right decision rapidly in the dynamic environment. These two challenges have not been solved and they restrict the development of efficient resource allocation methods in V2X communications.

  To deal with the discrete and continuous action space, a parameterized deep Q-network framework was proposed in \cite{xiong2018parametrized} to handle the hybrid action space based on the definition of hierarchical structure on the actions rather than using continuous actions discretization. Furthermore, the method in \cite{xiong2018parametrized} was adopted in \cite{wang2020drl} to solve the power and subchannel allocation of the nonorthogonal multiple access system by proposing a joint DRL algorithm. Although the proposed joint algorithm in \cite{wang2020drl} is able to efficiently handle the discrete-continuous hybrid action space in wireless communications systems, this algorithm will cause a high overhead and inaccurate resources allocation for V2X communications because it is designed based on a centralized setup in which the BS acts as the central controller to allocate resources for each user. Hence, it is necessary to design a distributed DRL algorithm, which is suitable for V2X communications to efficiently handle decision-making on the combination of power and spectrum resources. Following the idea in \cite{xiong2018parametrized} and \cite{wang2020drl}, we propose to use DQN to handle discrete sub-band assignment and deep deterministic policy gradient (DDPG) to handle continuous power allocation in order to overcome the first challenge caused by the hybrid action space of V2X communications. The ability of meta-learning in solving mismatch issues has been proven in computer vision \cite{finn2017model} and  wireless communications \cite{yuan2020transfer,park2019learning}. Hence, we propose to incorporate meta-learning into DRL to overcome the second challenge caused by the continuously changing environment in V2X communications. To be specific, we first propose a distributed DRL-based algorithm to achieve the optimal  solution of sub-band assignment and power allocation for V2X communications in which V2V links need to share the sub-band resources with V2I links. Then, we propose a meta-based DRL algorithm to improve the adaptation ability in the dynamic environment. We summarize the main contributions of this paper as follows:
\begin{itemize}
\item We propose a joint DRL-based algorithm to simultaneously improve the performance of V2I and V2V links in V2X communications in which the preassigned sub-bands to the V2I links need to be shared by V2V links. This algorithm uses DQN to solve the discrete sub-band assignment issue and employs DDPG to solve the continuous power allocation issue.
\item We propose a meta-based DRL algorithm by incorporating the idea of
Model-Agnostic Meta-Learning (MAML) \cite{finn2017model} into DRL. Different from the on-policy MAML, our algorithm focuses on solving the off-policy problem. Two-level update procedures are introduced in the meta-training stage of the proposed algorithm, which aim  to train a policy with   good generalization. The meta-adaptation stage is used to quickly adapt the trained policy in the new environment via a few time steps.
\item Extensive simulations are provided to evaluate the resource allocation performance of the proposed joint-DRL algorithm and the generalization capability of the proposed meta-based DRL algorithm in three realistic V2X communications scenarios. The results demonstrate that the proposed algorithms can efficiently solve the continuous power allocation issue and can achieve fast adaptation in the new environments.
\end{itemize}

The remainder of this paper is organized as follows. The system model and problem formulation are introduced in Section \ref{system_model}. Full details of the proposed joint DRL-based algorithm are presented in Section \ref{RL_algorithm}. Section \ref{meta_based_algorithm} describes the details of the proposed meta-based DRL algorithm. Simulation results and conclusions are presented in Section \ref{simu} and Section \ref{conc}, respectively.

{\em Notations:} The boldface lower case letter is used to represent a column vector. The notation $[A]_a^{b}$ denotes the value of $A$ that is lower bounded by $a$ and upper bounded by $b$. $\leftarrow$ denotes the assignment operation.  $\mathcal{N}(\mu, \eta^2)$ denotes a normal distribution with mean $\mu$ and variance $\eta^2$.

\section{System Model and Problem Formulation}\label{system_model}
\begin{figure}[ht]
\centering
\includegraphics[width=3.4in]{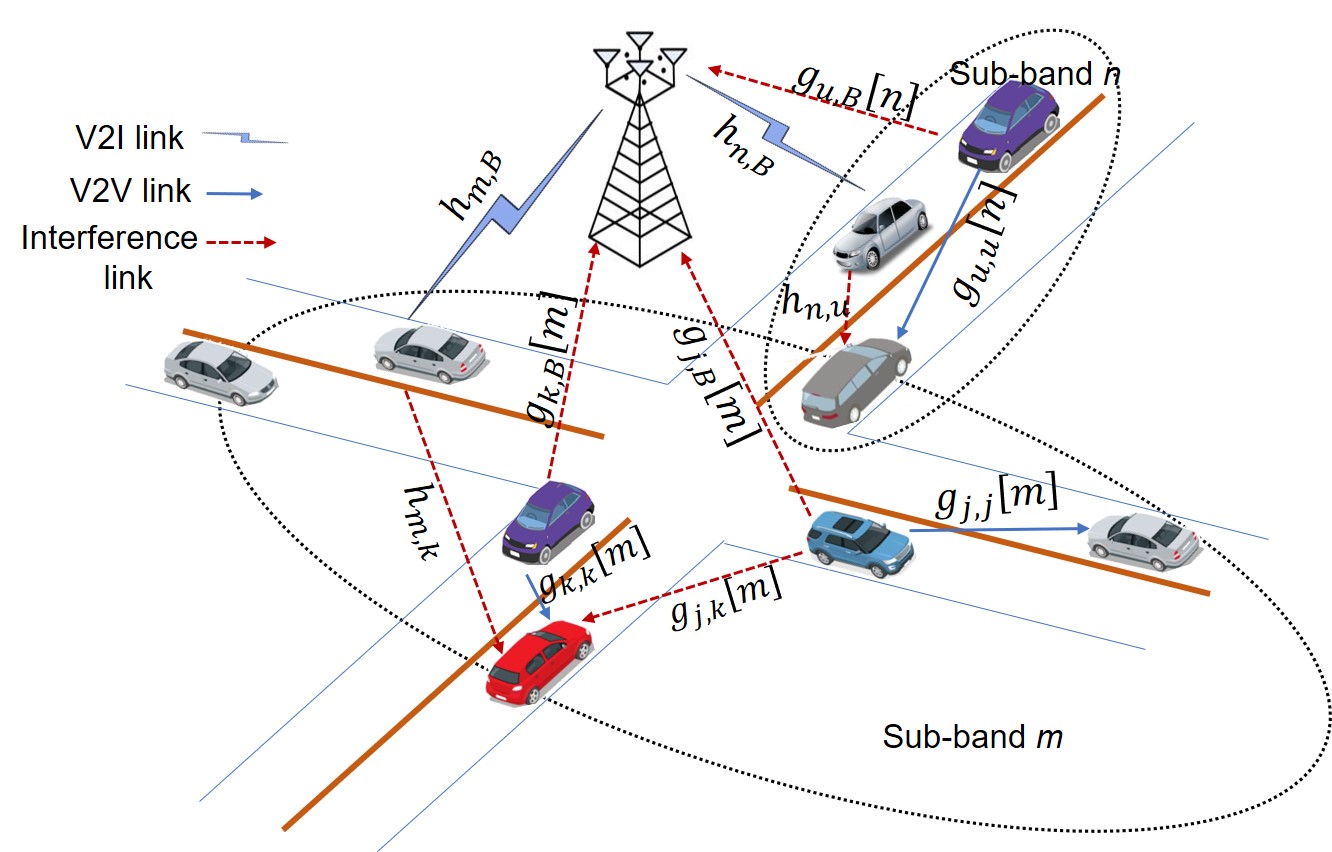}
\caption{A V2X communications scenario including $M$ V2I links and $K$ V2V links.}
\label{system}
\end{figure}
We consider a single-cell V2X communications network which includes one single-antenna base station (BS) and multiple single-antenna vehicular users, as shown in Fig. \ref{system}. Based on different service requirements in V2X communications \cite{3GPPTR37885}, all vehicles are divided into two groups: $M$ V2I links and $K$ V2V links. In this paper, we consider the  uplink of V2I communications. Specifically, the V2I links are used to upload high data rate information from vehicles to the BS, and the V2V links are used to deliver reliable safety-critical messages among vehicles. In our work, Mode 4 defined in the cellular V2X architecture is used as a distributed mechanism for spectrum selection of V2V links. Based on Mode 4, each vehicle can autonomously select radio resources for its V2V link rather than depending on the BS to allocate resources \cite{molina2017lte}. We assume that the number of sub-bands equals to the number of V2I links $M$ and each V2I link is preassigned with one orthogonal sub-band with fixed transmission power. In order to improve the spectrum utilization efficiency, the $M$ sub-bands allocated for V2I links are shared by V2V links. In addition, each V2V pair can only select one sub-band for their communications. Each sub-band can be shared by multiple V2V pairs.

In our paper, we assume that perfect CSI is available at each transmitter. This ideal assumption is possible since each transmitter can estimate the channel gain either in the uplink or through feedback from the users. The channel power gain is considered in this paper which includes the large-scale fading component and the small-scale fading component.  We express channel gain as $h=\alpha\tilde{h}$, where $\tilde{h}$ and $\alpha$ denote the small-scale fading and the large-scale fading including the path loss and shadowing for each communications link, respectively. Based on the channel guideline for V2X communications in the 3GPP release 14 \cite{3gpp2017technical}, we assume that the small-scale fading follows the Rayleigh distribution with zero mean and unit variance. In this paper, we define the channel gains for the $m$-th V2I link and the $k$-th V2V link over the sub-band $m$ as $h_{m,B}$ and $g_{k,k}[m]$, respectively. The interfering channel gain for the $m$-th V2I link from the $k$-th V2V link over the $m$-th sub-band is $g_{k,B}$[m]. The interfering channel gains received at the receiver of the $k$-th V2V pair from the transmitter of the $m$-th V2I link and the $j$-th V2V pair over the $m$-th sub-band are given by $h_{m,k}$ and $g_{j,k}$, respectively. The received signal to interference plus noise (SINR) ratio for the $m$-th uplink V2I link and for the $k$-th V2V link over the $m$-th sub-band are given by, respectively,
\begin{align}
\gamma_m^{v2i} &= \frac{p_m^{v2i}h_{m,B}}{\sum_{k=1}^K\rho_{k}[m]p_k^{v2v}[m]g_{k,B}[m]+\sigma^2},\label{sinr_v2i}\\
\gamma_k^{v2v}[m] &= \frac{p_k^{v2v}[m]g_{k,k}[m]}{I_k[m]+\sigma^2},\label{sinr_v2v}
\end{align}where $I_k[m]=p_m^{v2i}h_{m,k}+\sum_{k\neq j}\rho_{j}[m]p_j^{v2v}[m]g_{j,k}[m]$,  the first and second terms of $I_k[m]$ represent the interference received at the receiver of the $k$-th V2V pair from the $m$-th V2I link and the other V2V links that share the $m$-th sub-band,  $\sigma^2$ denotes the noise power for both links, $p_m^{v2i}$ and $p_k^{v2v}[m]$ denote the transmission power of the $m$-th V2I link transmitter and the $k$-th V2V link transmitter over the $m$-th sub-band, respectively. The binary variable $\rho_{k}[m]\in\{0,1\}$ denotes the sub-band selection indicator that $\rho_{k}[m]=1$ if the $k$-th V2V link uses the $m$-th sub-band, otherwise, $\rho_{k}[m]=0$. As we assume that each V2V link can only use one sub-band at the same time, $\rho_{k}[m]$ can be constrained as $\sum_{m=1}^M\rho_{k}[m]\leq1$. Then, the achievable data rate of the $m$-th V2I link and the $k$-th V2V link can be expressed as, respectively,
\begin{align}
R_m^{v2i} &= W\log(1+\gamma_m^{v2i}),\label{rate_v2i}\\
R_k^{v2v} &= \sum_{m=1}^M\rho_{k}[m]W\log(1+\gamma_k^{v2v}[m]),\label{rate_v2v}
\end{align}where $W$ denotes the bandwidth for each sub-band.

As mentioned earlier, the V2I links and V2V links need to satisfy different service requirements during the communications stages. \textcolor{black}{The V2I links aim to provide high quality entertainment services, which can be expressed as a maximization of their sum rate $\sum_{m=1}^MR_m^{v2i}$. The V2V links mainly focus on the reliable transmission of safety-critical messages, which aims to achieve high successful transmission probability of all V2V users during the information transmission period. It is more involved to derive the objective function of V2V links compared to V2I links. To this end, we first define the  successful transmission for each V2V link via the following condition
\begin{align}\label{condition}
\sum_{t=t_k}^{T/\Delta_T+t_k}\Delta_TR_k^{v2v}(t)\geq B, k=1,\ldots,K,
\end{align}where $B$ is the size of the payload for each V2V link, $T$ and $\Delta_T$ denote the maximum delay tolerant and the duration of each time slot, respectively, $t$ is the time slot index corresponding to the transmission time, and $t_k$ is the starting time of the $k$-th V2V link to transmit the payload. Note that the starting time of each V2V link to transmit its payload may not be the same due
to the asynchronous communications considered in this paper. According to the condition in \eqref{condition}, the message delivery of a V2V link is successful if the duration of delivering the payload $B$ does not exceed the maximum delay tolerant $T$. Otherwise the delivery is unsuccessful. We use $\omega_{k,u}$ as an indicator of the successful transmission of the $k$-th V2V link at its $u$-th payload transmission, i.e., $\omega_{k,u}=1$ if the transmission is successful, otherwise, $\omega_{k,u}=0$. Therefore, the successful transmission probability of all V2V links during a given transmission period can be expressed as
\begin{align}
\frac{\sum_{k=1}^K\sum_{u=1}^{O_k}\omega_{k,u}}{\sum_{k=1}^{K}O_k},
\end{align}where $O_k$ is the times of  payload transmission of the $k$-th V2V link during the transmission period. Note that the times of payload transmission during the transmission period for each V2V link may be different since V2V links may spend less time than $T$ to finish the transmission if the payload has been transmitted. Thus, the resource allocation problem for the designed V2X communications system can be formulated as
\begin{subequations}\label{problem}
\begin{align}
&\max_{\mathbf{\rho},\mathbf{P}^{v2v}}~\left(\sum_{m=1}^MR_m^{v2i}, \left\{\frac{\sum_{k=1}^K\sum_{u=1}^{O_k}\omega_{k,u}}{\sum_{k=1}^{K}O_k}\right\}\right),\label{ob}\\
&~~~\mathrm{s.t.}~\sum_{m=1}^M\rho_k[m]\leq1,\label{c1}\\
&~~~~~~~~0\leq p_k^{v2v}[m]\leq P_{max},\forall k,m,\label{c3}
\end{align}
\end{subequations}}where $\mathbf{\rho}=\{\rho_1[1],\ldots,\rho_k[m],\ldots,\rho_K[M]\}$ and $\mathbf{P}^{v2v}=\{p_1^{v2v}[1],\ldots,p_k^{v2v}[m],\ldots,p_K^{v2v}[M]\}$ are the set of the sub-band selection indicators and the power allocations, respectively.  The resource allocation problem in \eqref{problem} is a multi-objective optimization problem that aims to simultaneously maximize the sum rate of V2I links and reliable payload delivery of V2V links.  This problem is NP-hard and involves a sequential decision to be made over multiple transmission time slots. Thus, it is difficult to solve using conventional model-based optimization methods. Hence, we propose to use DRL methods to deal with this specific multi-objective problem. Existing DRL methods cannot deal with the continuous power constraint \eqref{c3} and the mismatch issue. To tackle these challenges in solving the problem \eqref{problem} in the dynamic environment, we propose new DRL algorithms to design a decentralized algorithm in the following two sections, when the environment is stationary and dynamic, respectively.

\section{DRL-based Resource Allocation Algorithm}\label{RL_algorithm}
In this section, we aim to design an  advanced DRL algorithm to solve the resource management problem in \eqref{problem} in a stationary environment. Since each V2I link is preassigned a sub-band with the fixed transmit power, we focus on solving the sub-band assignment and the power allocation for V2V links. DRL is an important branch of machine learning methods, which uses deep neural network (DNN) to enhance the learning efficiency of reinforcement learning (RL) \cite{mnih2015human}. Full details of the proposed algorithm are presented below.

\begin{figure}[ht]
\centering
\includegraphics[width=3.5in]{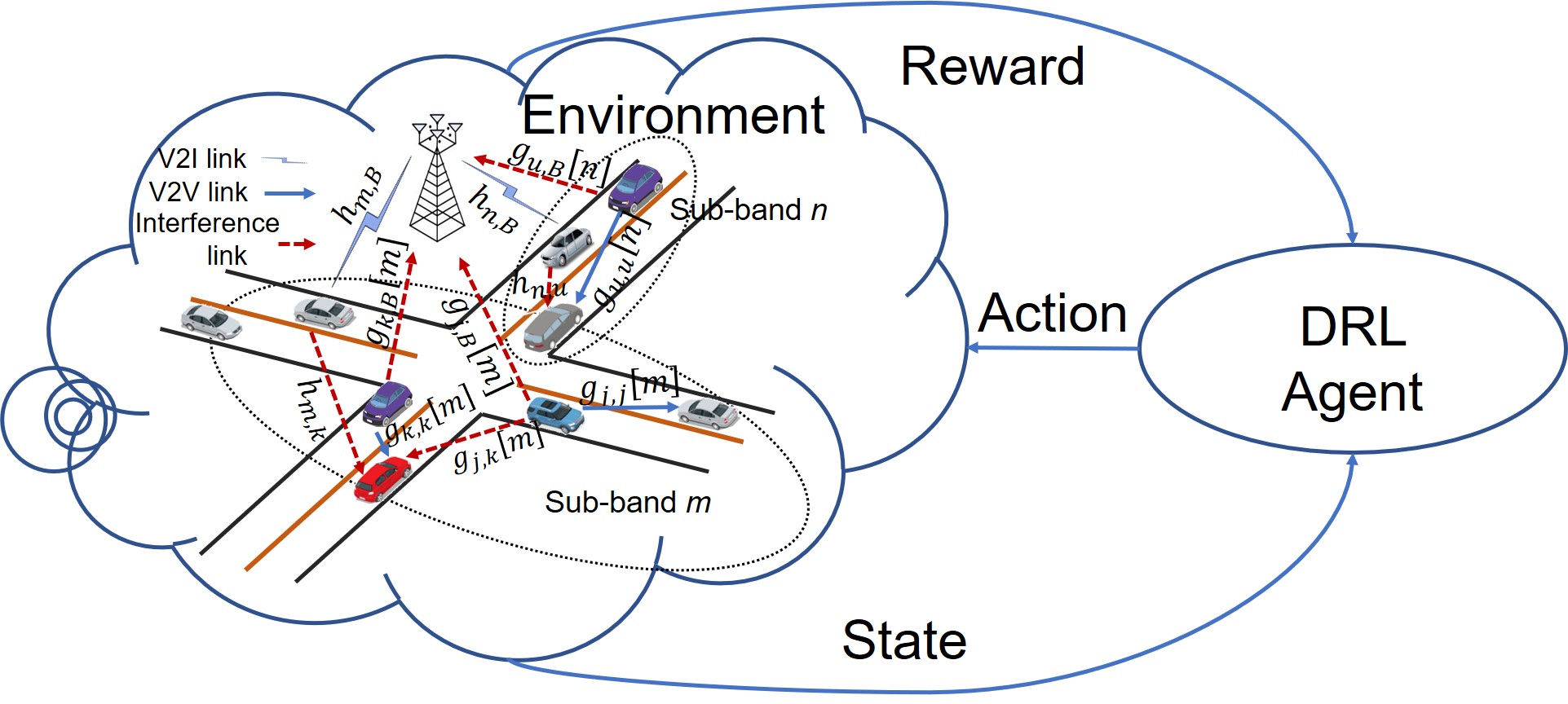}
\caption{The basic DRL architecture for V2X communications \cite{ye2019deep}.}
\label{mdp}
\end{figure}
DRL aims to solve an MDP by taking suitable actions through an agent by interacting with the unknown environment to maximize the reward. In order to use DRL to solve the problem in \eqref{problem}, we model our resource management problem as an MDP \cite{puterman2014markov}, which includes the DRL agent and the interactive environment shown in Fig. \ref{mdp}. Each V2V pair acts as an intelligent agent to make its own decision on the sub-band assignment and power allocation.  All V2V links make their own decisions at each time slot. Since the goal of DRL is to learn the best policy via maximizing the total accumulated reward, three key elements, i.e., state, action, and rewards need to be first defined for our problem \cite{luong2019applications}.
\subsection{Key Elements for MDP}
\textbf{State: }The environment state is an important part for policy learning since it includes some useful information such as driving state and channel information. We denote the state space as $\mathcal{S}$, which includes the states of all agents for each time slot.  The state $\mathbf{s}_t$ of a V2V agent at each time $t$ includes channel information, the received interference, the number of selected sub-bands for neighbors, the remaining load, and the remaining time. Specifically, the channel information for the $k$-th V2V agent over the $m$-th sub-band at time slot $t$ can be expressed as $\mathbf{G}_k^t[m]=\{g_{k,k}^t[m],h_{m,k}^t,g_{j,k}^t[m],g_{k,B}^t[m]\}$,  which includes the instantaneous channel gain of its own link over the $m$-th sub-band, $g_{k,k}^t[m]$, the interference channel gain from the transmitter of the $m$-th V2I link and the $j$-th V2V link $j\neq k$ over the $m$-th sub-band, $h_{m,k}^t$ and $g_{j,k}^t[m]$, and the interference channel gain from its transmitter to the BS, $g_{k,B}^t[m]$. The received interference power at the receiver of the $k$-th V2V over the $m$-th sub-band at the previous time slot is denoted as $\mathbf{I}_k^{t-1}[m]$. We use $\mathbf{N}_k^{t-1}[m]$ to present the number of times of the selected $m$-th sub-band occupied by the neighbors at the previous time slot. The remaining load and the remaining time used to meet the latency constraint are defined as $L_k^t$ and $U_k^t$, respectively. Thus, the environment state at time slot $t$ for the $k$-th V2V agent is given by
\begin{align}
&\mathbf{s}_k(t)=\{\{\mathbf{G}_k^t[m]\}_{m\in M}, \{\mathbf{I}_k^{t-1}[m]\}_{m\in M},\{\mathbf{N}_k^{t-1}[m]\}_{m\in M},\nonumber\\
&~~~~~~~~~~~L_k^t,U_k^t\}.
\end{align}

\textbf{Action:} Based on the observed state and policy, each V2V agent will make its own decision on the sub-band selection $\rho_k[m]$ and transmission power allocation $p_{k}^{v2v}[m]$ one by one, $k\in K,m\in M$.  We define the action space for all V2V agents as $\mathcal{A}=\{\mathcal{A}_k\}_{k=1}^K$, where $\mathcal{A}_k=\{\mathbf{a}_k^s,\mathbf{a}_k^p\}$ is the action space for the $k$-th V2V agent. $\mathbf{a}_k^s$ and $\mathbf{a}_k^p$ denote the set of possible sub-band assignment and power allocation decisions for the V2V agent $k$, respectively.  As mentioned, $M$ orthogonal sub-bands are preoccupied by $M$ V2I links and all V2V links will share these sub-bands, thus the set of possible sub-band assignment decisions for each agent at time slot $t$ can be defined as
\begin{align}
\mathbf{a}_k^s(t)=\{\rho_k[1](t),\ldots,\rho_k[M](t)\}, \forall k.
\end{align}The dimension of $\mathbf{a}_k^s$ is $M$. Similarly, the set of possible power allocation decisions can be defined as $\mathbf{a}_k^p(t)=\{p_{k}^{v2v}[1](t),\ldots,p_{k}^{v2v}[M](t)\},\forall k$ with dimension $M$. Since we have assumed that each V2V link can only use one sub-band at the same time, which means $p_{k}^{v2v}[m]=0$ if $\rho_k[m]=0$, and the set of power allocation decisions can be reformulated as $a_k^p(t)=\{p_{k}^{v2v}[m](t)\}_{m\in M}$ with dimension $1$. Therefore the dimension of the action space for each agent equals $M+1$.

\textbf{Rewards:} One of the advantages of RL for solving decision-making problems is that it can design a flexible reward to represent the hard-to-optimize multiple objectives and constraints.  The immediate reward will be returned by the environment once the agent takes an action based on the policy and observed the state. It indicates that the reward can reflect the performance of the decision made by the proposed policy. For our problem, a good decision on sub-band assignment and power allocation for each V2V link can maximize the sum rate of V2I links while improving the success probability of each V2V link to transmit payload within a certain time as much as possible. In order to reflect the performance of the decision taken by the agent, we consider three parts to formulate the immediate reward, which includes the sum rate of V2I links, the sum rate of V2V links, and the time used for transmission. Therefore, the immediate reward at time slot $t$ can be expressed as
\begin{align}\label{reward}
r_t=\nu^i\sum_{m=1}^MR_m^{v2i}+\nu^v\sum_{k=1}^KR_k^{v2v}-\lambda(T-U_k^t),
\end{align}where $\nu^i$, $\nu^v$, and $\lambda$ denote the positive weights of each part, and $T$ is the maximum tolerable latency. The expression $(T-U_k^t)$ denotes the time used for transmission, which can be considered as a penalty function. If $(T-U_k^t)$ increases, the remaining time will decrease, which means the  probability of successful payload delivery within the certain time limit will decrease. Together with the sum rate of V2V links, this expression will reflect the second objective in \eqref{problem}.  Since RL aims to find an optimal policy that can achieve the expected reward from the state in the long-term, the cumulative discounted reward can be defined as
\begin{align}\label{longreward}
R_t=\sum_{i=0}^{\infty}\gamma_ir_{t+i},
\end{align}where $\gamma_i\in[0,1]$ denotes the discount factor, which is used to balance the future reward and the current reward. The cumulative reward equals the immediate reward when $\gamma_i=0$.
\subsection{DRL-based Decentralized Algorithm}
\begin{figure}[ht]
\centering
\includegraphics[width=3.5in]{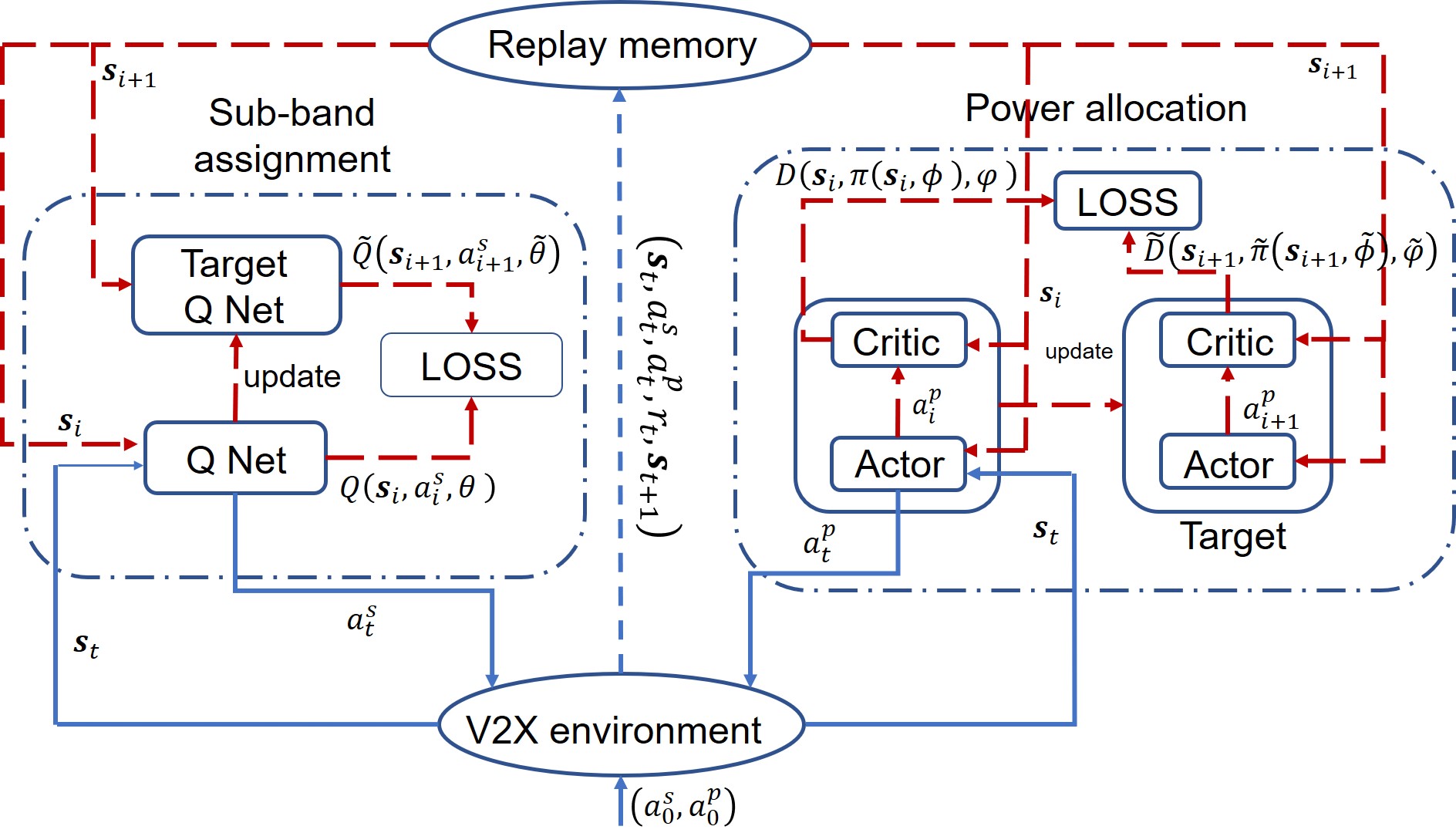}
\caption{The architecture of the joint DRL framework.}
\label{joint_model}
\end{figure}
Based on the definition of the key elements in the above subsection, we can introduce the proposed decentralized algorithm. The fundamental concept of DRL is to design an optimal policy for the agent to achieve the optimal mapping from the state to the action. Hence, how to achieve this mapping is the main part of our algorithm. As mentioned before, each agent needs to take two actions based on the observed state, one is for sub-band assignment and the other one is for power allocation. As the action space of our work includes both discrete and continuous actions, the DQN method, which has a discrete output, cannot be directly used to take the action from the action space. To solve the challenge on the joint action space, existing works first quantize the continuous transmission power into $L$-level discrete values, and then use DQN to design the algorithm. However, quantizing the continuous action not only causes the quantization error but also increases the dimension of the action space. Besides, how to provide an efficient quantization method on the transmit power is difficult since the quantization may lose some useful power information. In order to efficiently handle the problem, we use DQN to make the decision on the sub-band assignment and use DDPG to make the decision on the power allocation for each V2V agent. The designed framework is shown in Fig. \ref{joint_model}, which includes the decision making units for sub-band assignment and power allocation as well as the learning process for both units. Each V2V agent starts with the initial actions $(a_0^s,a_0^p)$ to interact with the V2X communications environment. Each V2V agent will receive its corresponding immediate reward and new state from the environment after taking its actions on the sub-band and power based on the current state. The decision making processes of each V2V agent for the sub-band and power allocation depend on the current state and policy. In the following, we introduce how the joint actions will be taken by each V2V agent at time slot $t$ (solid line in Fig. \ref{joint_model}) and how to update the policy for both units (dash line in Fig. \ref{joint_model}).
\subsubsection{DQN for Sub-band Assignment}
In this part, we describe how to use DQN to make the decision on the sub-band selection. Deep Q-learning (DQL), which is implemented by combining DNN and Q-learning, has been considered as one of the useful value-based off-policy DRL techniques on handling the problems with large state spaces and discrete actions \cite{mnih2015human}. The DNN used in deep Q-learning is called DQN, which is used to estimate the action-value function. The DQL technique can be directly applied to solve the sub-band assignment problem due to the discrete feature of the sub-band. The DQN unit is shown in the left part of Fig. \ref{joint_model}, which includes one $Q$ network and one target $Q$ network. The use of the target network is to improve the stability of DQL. Both networks have the same DNN architecture.  At the beginning of the time slot $t$, the $k$-th V2V agent chooses an action $a_k^s(t)$ from $\mathbf{a}_k^s(t)$ based on the $\epsilon$-greedy policy and the state $\mathbf{s}_k(t)$. The $\epsilon$-greedy policy is adopted to balance the exploration of new actions and the exploitation of known actions. It indicates that the $k$-th agent randomly selects $a_k^s(t)$ with probability $\epsilon\in(0,1)$, or selects the action $a_k^s(t)$ according to the following equation with probability $1-\epsilon$:
\begin{align}
a_k^s(t) = \mathrm{arg~max}_{a_k^s(t)\in\mathbf{\mathcal{A}}_k}[Q(\mathbf{s}_k(t),a_k^s(t),
\theta)],\forall k,
\end{align}where $Q(\mathbf{s}_k(t), a_k^s(t), \theta)$ is the output Q-value given the observed state $\mathbf{s}_k(t)$ and action $a_k^s(t)$ for the $Q$ network, and $\mathbf{\theta}$ denotes the weights of the $Q$ network.
\subsubsection{DDPG for Power Allocation}
In order to efficiently handle the continuous power action space, the DDPG unit is adopted to determine the power allocation. DDPG is an actor-critic off-policy RL algorithm \cite{lillicrap2015continuous}, which utilizes the advantages of DPG theorem \cite{silver2014deterministic} and the DQN algorithm. As shown in the right part of Fig. \ref{joint_model}, the DDPG unit uses the actor network to generate the deterministic action and uses the critic network to evaluate the reward of the state-action pair. Similar to DQN, the target network is also used for the actor and critic network in DDPG in order to improve the stability. After obtaining the selected sub-band for the $k$-th agent with the observed state $\mathbf{s}_k(t)$, DDPG is applied to allocate power for this agent on the corresponding sub-band. Specifically, the $k$-th agent uses the actor network to deterministically generate power allocation $a_k^p(t)$ via the same state $\mathbf{s}_k(t)$, $a_k^p(t) = \pi(\mathbf{s}_k(t),\mathbf{\phi})$, where $\pi(\mathbf{s}_k(t),\mathbf{\phi})$ denotes the policy of the actor network with the network weights $\mathbf{\phi}$. In order to balance the exploration and the exploitation, a stochastic noise is introduced \cite{lillicrap2015continuous}. Hence, the decision made by the actor network on power allocation for the agent $k$ is given by
\begin{align}\label{powerselect}
a_k^p(t) = [\pi(\mathbf{s}_k(t),\mathbf{\phi})+ \eta]_{0}^{P_{max}},\forall k
\end{align}where $\eta$ follows a normal distribution $\mathcal{N}(0,0.2)$. The lower and upper bounds $0$ and $P_{max}$ are used to enforce the power constraint.
\subsubsection{Network Update for Both Units}
After obtaining the actions $(a_k^s(t),a_k^p(t))$ based on the observed state $s_k(t)$ at time slot $t$, the immediate reward $r_k(t)$ and the new state $\mathbf{s}_k(t+1)$ will be returned to the agent $k$ from the environment. Afterwards, the experience $(\mathbf{s}_k^t,\mathbf{a}_k(t),r_k(t),\mathbf{s}_k(t+1))$ obtained at time slot $t$ for the agent $k$ is stored in the replay memory block $\mathcal{D}$ with size $U$ by using the experience replay strategy \cite{mnih2015human}. Then, a mini-batch of experiences $(\mathbf{s}_k(i),\mathbf{a}_k(i),r_k(i),\mathbf{s}_k(i+1))$ with size $N_{tr}$ is randomly selected by the agent $k$ from the replay memory. Notice that the selected experiences may include the experiences of different agents at  different time slots. The aim of using the mini-batch of experiences rather than the current time slot experience is to ensure that the data used for training is independently and identically distributed. The symbol $i$ in the bracket denotes the experience of the $i$-th time slot. Based on the states $\mathbf{s}_k(i)$ and $\mathbf{s}_k(i+1)$ selected in the mini-batch, the output $Q$-value at the $Q$ network and target $Q$ network can be expressed as $Q(\mathbf{s}_k(i),a_k^s(i),\mathbf{\theta})$ and $\tilde{Q}(\mathbf{s}_k(i+1), a_k^s(i+1),\tilde{\mathbf{\theta}})$, respectively, where $\tilde{\mathbf{\theta}}$ is the weights for the target $Q$ network. The difference of the output $Q$ value between $Q$ and target $Q$ networks can be measured by using the following loss function
\begin{align}\label{lossdqn}
L_k^{D}(\mathbf{\theta})&=\sum_i(r_k(i)+\max_{a_k^s(i+1)\in\mathbf{\mathcal{A}}_k}
\tilde{Q}(\mathbf{s}_k(i+1),a_k^s(i+1),\tilde{\mathbf{\theta}})\nonumber\\
&~~~~~~~~-Q(\mathbf{s}_k(i),a_k^s(i),\theta))^2,\forall k.
\end{align}Then, the weights $\mathbf{\theta}$ of the $Q$ network can be updated by minimizing the loss function \eqref{lossdqn} of the $k$-th agent using the gradient descent technique as follows,
\begin{align}
\mathbf{\theta}\leftarrow\mathbf{\theta}-\alpha\nabla_{\mathbf{\theta}}L_k(\mathbf{\theta}),
\end{align}where $\alpha$ is the learning rate. The weights variable $\tilde{\mathbf{\theta}}$ of the target network is periodically updated by $\mathbf{\theta}$.

Next, we move to updating the weights of the networks in DDPG based on the same selected mini-batch experiences. By using the selected experiences $(\mathbf{s}_k(i),\mathbf{a}_k(i),r_k(i),\mathbf{s}_k(i+1))$, the estimated $Q$ values of the critic network and the target critic network can be expressed as $Q(\mathbf{s}_k(i), a_k^p(i),\mathbf{\varphi})$ and $\tilde{Q}(\mathbf{s}_k(i+1),\tilde{\pi}(\mathbf{s}_k(i+1),\tilde{\mathbf{\phi}}),\tilde{\mathbf{\varphi}})$, where $\tilde{\pi}(\mathbf{s}_k(i+1),\tilde{\mathbf{\phi}})$ represents the target actor network with the weights   $\tilde{\mathbf{\phi}}$, and $\tilde{\mathbf{\varphi}}$ is the weights of the target critic network. The difference between the critic and target critic networks can be represented by using the following loss function
\begin{align}\label{lossrbddpg}
L_k^{C}(\mathbf{\varphi}) = \sum_{i}(y_k(i)-Q(\mathbf{s}_k(i),a_k^p(i),\mathbf{\varphi}))^2,\forall k,
\end{align}where $y_k(i)=r_k(i)+\gamma\tilde{Q}(\mathbf{s}_k(i+1),\tilde{\pi}(\mathbf{s}_k(i+1),\tilde{\mathbf{\phi}}),
\tilde{\mathbf{\varphi}})$, $\mathbf{\varphi}$ is the weights of the critic network. Minimization of \eqref{lossrbddpg} can be solved by using the gradient descent technique as $\mathbf{\varphi}\leftarrow\mathbf{\varphi}-\beta_c\nabla_{\mathbf{\varphi}}L_k(\mathbf{\varphi})$, where $\beta_c$ is the learning rate. According to the DPG theorem \cite{silver2014deterministic}, the actor network updates its weights in the direction of getting larger cumulative discounted reward. Thus, the weights of the actor network $\mathbf{\phi}$ can be updated by minimizing the following loss function
\begin{align}\label{losspowerddpg}
L_k^{A}(\mathbf{\phi}) = -Q(\mathbf{s}_k(i), \pi(\mathbf{s}_k(i), \mathbf{\phi}),\mathbf{\varphi}),\forall k.
\end{align}The weights of the target networks for actor and critic in DDPG are updated by the following equations
\begin{align}
\tilde{\mathbf{\phi}}&\leftarrow \tau\mathbf{\phi}+(1-\tau)\tilde{\mathbf{\phi}},\label{actortarget}\\
\tilde{\mathbf{\varphi}}&\leftarrow \tau\mathbf{\varphi}+(1-\tau)\tilde{\mathbf{\varphi}},\label{cristictarget}
\end{align}where $\tau\in[0,1]$ is the update frequency factor used to control the fraction of the weights of the main network to copy to the target network.

After all agents update the weights of the networks in DQN and DDPG at time slot $t$, the algorithm will move to the next time slot to generate the experience of each agent and to update the weights  of networks based on the sampled mini-batch experiences. Full details of the proposed decentralized DRL-based algorithm are summarized in Algorithm 1. Note that the convergence of Algorithm 1 depends on the offline training process since it aims to train   efficient neural networks for the DQN and DDPG units.   Algorithm 1 also  converges when the training process is finished.
\begin{table}\label{Algorithm1}
\hrule
\vspace{1mm}
\noindent \textbf{Algorithm 1:}  Combined DQN and DDPG Based Resource Management Algorithm \label{Table: Table II}
\vspace{1mm}
\hrule
\vspace{1mm}
\begin{enumerate}
\item Initialize the weights $\mathbf{\theta}$ of DQN unit, the weights $\mathbf{\phi}$ and $\mathbf{\varphi}$ of DDPG unit.
\item Initialize the weights of target network for DQN and DDPG as $\tilde{\mathbf{\theta}}=\mathbf{\theta}$, $\tilde{\mathbf{\phi}}=\mathbf{\phi}$, and $\tilde{\mathbf{\varphi}}=\mathbf{\varphi}$.
\item \textbf{for} each step $t$ \textbf{do}
\item $\hspace*{3mm}$ Initialize the V2I and V2V communications scenario.
\item $\hspace*{3mm}$ \textbf{for} each V2V agent $k$ \textbf{do}
\item $\hspace*{6mm}$ Observe the state $\mathbf{s}_k(t)$.
\item $\hspace*{6mm}$ Choose sub-band action $a_k^s(t)$ following the $\epsilon$-greedy policy based on $\mathbf{s}_k^t(t)$.
\item $\hspace*{6mm}$ Generate power action $a_k^p(t)$ by \eqref{powerselect}.
\item $\hspace*{6mm}$ Evaluate the immediate reward $r_k(t)$ in \eqref{reward} and next state $\mathbf{s}_k(t+1)$ by executing the actions $(a_k^s(t), a_k^p(t))$.
\item $\hspace*{6mm}$ Store the experience $(\mathbf{s}_k(t),a_k^s(t), a_k^p(t),r_k(t),\mathbf{s}_{k+1}(t))$ into the memory block $\mathcal{D}$.
\item $\hspace*{6mm}$ Randomly sample a mini-batch of experiences from $\mathcal{D}$.
\item $\hspace*{6mm}$ Update weights $\mathbf{\theta}$, $\mathbf{\phi}$, and $\mathbf{\varphi}$ by minimizing the corresponding loss function in \eqref{lossdqn}, \eqref{lossrbddpg}, and \eqref{losspowerddpg} via the gradient\\ $\hspace*{7mm}$descent technique.
\item $\hspace*{6mm}$ Update the weights $\tilde{\mathbf{\phi}}$ and $\tilde{\mathbf{\varphi}}$ of target networks in DDPG by \eqref{actortarget} and \eqref{cristictarget}.
\item $\hspace*{6mm}$ Update the weights $\tilde{\mathbf{\theta}}$ of the target $Q$ network every $100$ steps by copying weights $\mathbf{\theta}$.
\item $\hspace*{3mm}$ \textbf{end for}
\item \textbf{end for}
\item Output the trained network weights $\mathbf{\theta}$, $\mathbf{\phi}$, and $\mathbf{\varphi}$.
\end{enumerate}
\hrule
\end{table}

\section{Meta-reinforcement Learning}\label{meta_based_algorithm}
 In Section \ref{RL_algorithm}, an efficient DRL algorithm has been proposed to solve the sub-band assignment and power allocation problem in a stationary V2X communications system. The algorithm is designed based on the assumption that the communications environment and QoS requirements remain the same during the training and testing stages. However, this assumption is   impractical in real-time V2X communications. Existing algorithms lack the adaptability  for different communications scenarios since there will be mismatch if the testing environment follows a different distribution from the training environment. To address this challenge, we propose a meta reinforcement learning algorithm, which can provide a good reinforcement learning model that has the generalization ability to new environments and new tasks.  The meta reinforcement learning algorithm is designed by incorporating the idea of the MAML algorithm \cite{finn2017model} into the proposed joint RL framework. The MAML-based reinforcement learning proposed in \cite{finn2017model} cannot be directly used to solve our problem since it focuses on solving the on-policy MDP. The off-policy MDP is considered in our resource allocation problem. In the following, we describe the proposed meta reinforcement learning algorithm for our problem based on the idea of the MAML framework.
\subsection{Definitions}
The aim of meta-learning is to train a reinforcement model, which can fast adapt to the new tasks. It indicates that each V2V agent needs to learn a variety of different tasks. In order to achieve this goal, we first define a meta task set $\mathcal{T}$ that includes $N_T$ tasks. Each task $\mathcal{T}_{j} (j=1,\ldots,N_T)$ is considered as an MDP $(\mathcal{S},\mathcal{A},\mathcal{R},\tilde{\mathcal{S}})$ \textcolor{black}{to train the optimal policy of resource allocation in the environment where  each vehicle has different initial positions}. Each task contains state $\mathcal{S}$, action $\mathcal{A}$, reward $\mathcal{R}$, and new state $\tilde{\mathcal{S}}$. We define a replay buffer for task $\mathcal{T}_{j}$ as $D_{\mathcal{T}_{j}}$, which is used to store the experiences. The support set and the query set of each task are defined as $\mathbb{D}_j^{tr}$ and $\mathbb{D}_j^{val}$, respectively. The support set and query set are used for the weights updating of the global network  and individual task networks in the meta-training stage, respectively.
\subsection{Meta-training Stage}
Meta-training is an important part of the meta-learning algorithm since it aims to train initialized parameters  for the neural network that can fast adapt to a new task. Based on the MAML algorithm on training the parameters initialization, we employ a two-level training mechanism  to design the meta reinforcement learning algorithm: one is called individual-level update and the other is called global-level update. The former is a step-by-step optimization process on each task and the latter is a periodic synchronous updating process on a batch of sampled tasks. Each task performs individual-level update on its own parameter based on the inherited globally-shared initialization of parameters, then contributes to the global parameters update based on its own parameter. The training process of both   updates uses the same neural network architecture. We provide details of these two update steps below.

The individual-level update can be considered as a process of learning the policy on selecting sub-band and power for each task based on the sampled mini-batch experiences. This process aims to optimize the parameters of the constructed three networks (DQN, actor, and critic) for each task via the globally-shared initialization of parameters. In Section \ref{RL_algorithm}, we have defined the function and construction of three networks for an MDP task. Thus, all equations used to update the network weights in Section \ref{RL_algorithm} can be directly utilized in meta learning. Note that each task uses the same optimization problems to obtain its own weights of networks via sampling the different experiences from the replay memory. Therefore, the weights of three networks of each task can be optimized by using the following optimization problems, which are specific to task $j$:
\begin{align}\label{taskoptim}
\left\{
\begin{aligned}
\hat{\mathbf{\theta}}_j&=\mathrm{arg~min}_{\mathbf{\theta}}
L_k^{D}(\mathbf{\theta},\mathbb{D}_{j}^{tr}),\forall_k,\\
\hat{\mathbf{\phi}}_j&=\mathrm{arg~min}_{\mathbf{\phi}}
L_k^{A}(\mathbf{\phi},\mathbb{D}_{j}^{tr}),\forall_k,\\
\hat{\mathbf{\varphi}}_j&=\mathrm{arg~min}_{\mathbf{\varphi}}
L_k^{C}(\mathbf{\varphi},\mathbb{D}_{j}^{tr}),\forall_k,
\end{aligned}
\right.\!\!\!\!
\end{align}where $\hat{\mathbf{\theta}}_j$, $\hat{\mathbf{\phi}}_j$, and $\hat{\mathbf{\varphi}}_j$ denote the network weights of the $Q$ network, actor network, and critic network, respectively, $L_k^{D}(\hat{\mathbf{\theta}}_j,\mathbb{D}_{j}^{tr})$ is the loss function of the DQN at the agent $k$ which is defined in \eqref{lossdqn}. Similarly, the definitions of the loss functions $L_k^{C}(\mathbf{\varphi}_j,\mathbb{D}_{j}^{tr})$, $L_k^{A}(\mathbf{\phi}_j,\mathbb{D}_{j}^{tr})$ for the critic and actor networks can be found in \eqref{lossrbddpg} and \eqref{losspowerddpg}, respectively. Notice that the weights of each network of each task needs to be updated via all agents. $\mathbb{D}_{j}^{tr}$ is the set of the sampled experiences from the memory buffer $D_{\mathcal{T}_{j}}$ of task $j$. Since the loss function in \eqref{taskoptim} for each network of each task is differentiable, the gradient descent method can be used to update the network weights in \eqref{taskoptim} based on the sampled experiences. Notice that the parameter updating process for each task is independent. \textcolor{black}{Based on the multiple gradient updates, the parameters of the network in DQN and DDPG for task $j$ can be updated by using the following equations
\begin{align}\label{innermore}
\left\{
\begin{aligned}
\hat{\mathbf{\theta}}_j^{(n)}&=\hat{\mathbf{\theta}}_j^{(n-1)}-
\hat{\alpha}\nabla_{\hat{\mathbf{\theta}}_j^{(n-1)}}L_k^{D}(\hat{\mathbf{\theta}}_j^{(n-1)},\mathbb{D}_j^{tr}),\forall k,\\
\hat{\mathbf{\varphi}}_j^{(n)}&=\hat{\mathbf{\varphi}}_j^{(n-1)}-\hat{\beta}_{c}
\nabla_{\hat{\mathbf{\varphi}}_j^{(n-1)}}L_k^{C}(\hat{\mathbf{\varphi}}_j^{(n-1)},\mathbb{D}_j^{tr}),\forall k,\\
\hat{\mathbf{\phi}}_j^{(n)}&=\hat{\mathbf{\phi}}_j^{(n-1)}+\hat{\beta}_{a}
\nabla_{\hat{\mathbf{\phi}}_j^{(n-1)}}L_k^{A}(\hat{\mathbf{\phi}}_j^{(n-1)},\mathbb{D}_j^{tr}),\forall k,
\end{aligned}
\right.\!\!\!\!
\end{align}
where $\hat{\alpha}$, $\hat{\beta}_{c}$, and $\hat{\beta}_{a}$ denote the learning rate of individual-level update for the $Q$ network, the critic network, and the actor network, respectively, and the superscript $n$ denotes the index of the iteration. Note that network parameters of each task are updated by the corresponding global network parameters at the first iteration step $(n=1)$, which indicates $\hat{\mathbf{\theta}}_k^{(0)}=\mathbf{\theta}$, $\hat{\mathbf{\varphi}}_j^{(0)}=\mathbf{\varphi}$, $\hat{\mathbf{\phi}}_j^{(0)}=\mathbf{\phi}$, and then updated by its own parameters obtained at the previous iterative step.} After all tasks in the batch finish the updating of their own network parameters, the global parameters can be updated based on these task parameters as described below.

Global-level update is an updating process of the global network parameters. This process is achieved by aggregating the adaptation ability of the trained policy for each task on their new sampled experiences. When every task in the batch finishes its own network parameters updating, the adaptation ability of the updated policy for each task can be evaluated by estimating the loss function over its corresponding query set $\mathbb{D}_j^{val}$. By adding such loss functions together, the loss function used to optimize the global network parameters $(\mathbf{\theta},\mathbf{\varphi},\mathbf{\phi})$ can be formed as $\sum_{j}L_j^{D}(\hat{\mathbf{\theta}}_j,\mathbb{D}_j^{val})$, $\sum_{j}L_j^{C}(\hat{\mathbf{\varphi}}_j,\mathbb{D}_j^{val})$, and $\sum_{j}L_j^{A}(\hat{\mathbf{\phi}}_j,\mathbb{D}_j^{val})$, respectively. {Thus, the optimization problems used to optimize $\mathbf{\theta}$, $\mathbf{\varphi}$, and $\mathbf{\phi}$ can be expressed as, respectively,
\begin{align}\label{globalopti}
\left\{
\begin{aligned}
\mathbf{\theta}&=\mathrm{arg~min}_{\mathbf{\theta}}
\sum_{j}L_j^{D}(\hat{\mathbf{\theta}}_j,\mathbb{D}_{j}^{val}),\\
\mathbf{\mathbf{\varphi}}&=\mathrm{arg~min}_{\mathbf{\mathbf{\varphi}}}
\sum_{j}L_j^{C}(\hat{\mathbf{\mathbf{\varphi}}}_j,\mathbb{D}_{j}^{val}),\\
\mathbf{\phi}&=\mathrm{arg~min}_{\mathbf{\phi}}
\sum_{j}L_j^{A}(\hat{\mathbf{\phi}}_j,\mathbb{D}_{j}^{val}).
\end{aligned}
\right.\!\!\!\!
\end{align}Based on the gradient descent method, the parameters listed in \eqref{globalopti} can be updated by
\begin{align}\label{globalupdate}
\left\{
\begin{aligned}
\mathbf{\theta}&\leftarrow\mathbf{\theta}-\alpha\nabla_{\mathbf{\theta}}\sum_{j}
L_j^{D}(\hat{\mathbf{\theta}}_j,\mathbb{D}_j^{val}),\\
\mathbf{\varphi}&\leftarrow\mathbf{\varphi}-\beta_c\nabla_{\mathbf{\varphi}}\sum_{j}
L_j^{C}(\hat{\mathbf{\varphi}}_j,\mathbb{D}_j^{val}),\\
\mathbf{\phi}&\leftarrow\mathbf{\phi}+\beta_a\nabla_{\mathbf{\phi}}\sum_{j}
L_j^{A}(\hat{\mathbf{\phi}}_j,\mathbb{D}_j^{val}),
\end{aligned}
\right.\!\!\!\!
\end{align}where $\alpha$, $\beta_a$, and $\beta_c$ are the learning rate of the global-level update.} There exists a chain rule when updating the global parameters by using \eqref{globalupdate}. For instance, the gradient of the sum loss function over $\mathbf{\theta}$ needs to calculate the gradient of each task over its own parameter at   every iteration, that is $\frac{\partial L_j(\hat{\mathbf{\theta}}_j,\mathbb{D}_j^{val})}{\partial(\hat{\mathbf{\theta}}_j)}=\frac{\partial L_j(\hat{\mathbf{\theta}}_j^{G_{in}},\mathbb{D}_j^{val})}{\partial(\hat{\mathbf{\theta}}_j^{G_{in}})}
\cdot\frac{\partial(\hat{\mathbf{\theta}}_j^{G_{in}})}{\partial(\hat{\mathbf{\theta}}_j^{G_{in}-1})}\cdot
\frac{\partial(\hat{\mathbf{\theta}}_j^{G_{in}-1})}{\partial(\hat{\mathbf{\theta}}_j^{G_{in}-2})}
\cdot\ldots\cdot\frac{\partial(\hat{\mathbf{\theta}}_j^0)}{\partial\mathbf{\theta}}$. The chain rule for updating $\mathbf{\varphi}$ and $\mathbf{\phi}$ is similar to the update of $\mathbf{\theta}$. According to the calculation for the updating equation in \eqref{globalupdate}, the proposed meta DRL algorithm needs an  additional backward pass compared to the DRL algorithm proposed in Section \ref{RL_algorithm}. When the individual-level update and global-level update finish, the algorithm moves to the next batch to continuously update the global network parameters.

\subsection{Meta-adaptation Stage}
Meta-adaptation stage aims to adapt the trained parameters based on the generated experiences in the new environment. In the meta-training stage, we have learned the initial network parameters, which have good generalization ability. Based on the well trained parameters $\mathbf{\theta}$, $\mathbf{\varphi}$, and $\mathbf{\phi}$, the proposed meta-based DRL algorithm can achieve fast adaptation on the new task via a few steps. Similar to the parameter updating process in the individual-level update, the network parameters of the new task can be updated by
\begin{align}\label{adaptmore}
\left\{
\begin{aligned}
\hat{\mathbf{\theta}}&=\hat{\mathbf{\theta}}-\hat{\alpha}\nabla_{\hat{\mathbf{\theta}}}
L_k^{D}(\hat{\mathbf{\theta}}),\forall k,\\
\hat{\mathbf{\varphi}}&=\hat{\mathbf{\varphi}}-\hat{\beta}_{c}\nabla_{\hat{\mathbf{\varphi}}}
L_k^{C}(\hat{\mathbf{\varphi}}),\forall k\\
\hat{\mathbf{\phi}}&=\hat{\mathbf{\phi}}+\hat{\beta}_{a}\nabla_{\hat{\mathbf{\phi}}}
L_k^{A}(\hat{\mathbf{\phi}}),\forall k,
\end{aligned}
\right.\!\!\!\!
\end{align}where $\hat{\mathbf{\theta}}$, $\hat{\mathbf{\varphi}}$, and $\hat{\mathbf{\phi}}$ are initialized as the trained global parameters $\mathbf{\theta}$, $\mathbf{\varphi}$, and $\mathbf{\phi}$ at the beginning of the time step. The experiences are stored in the replay memory $\mathbb{D}_{ad}$. After adapting the parameters on the new task via the adaptation stage, the performance of the proposed meta-based DRL algorithm can be evaluated in the testing stage. Full details of meta-training and meta-adaptation are provided in Algorithm 2.  Similar to Algorithm 1, the convergence of Algorithm 2 depends on the offline training.
\begin{table}\label{Algorithm2}
\hrule
\vspace{1mm} 
\noindent \textbf{Algorithm 2:}  The proposed meta-based DRL. \label{Table: Table II}
\vspace{1mm}
\hrule
\vspace{1mm}
\textbf{Input:}  {Individual-level learning rate $(\hat{\alpha}$, $\hat{\beta}_c$, $\hat{\beta}_a)$, and global-level learning rate $(\alpha,\beta_c,\beta_a)$, task number $N_b$, the time step $T_{max}$, the number of individual-level iteration steps $G_{in}$, and the number of time slots in the adaptation stage $T_{Ap}$}
\vspace{1mm}
\hrule
\vspace{2mm}
$\hspace*{4mm}$$\mathbf{Meta-training}$
\vspace{-1mm}
\begin{enumerate}
\item Initialize the network parameter $\mathbf{\theta}$, $\mathbf{\varphi}$, and $\mathbf{\phi}$
\item \textbf{for} each batch \textbf{do}
\item $\hspace*{3mm}$ Sample $N_b$ tasks from the task set $\{\mathcal{T}\}$
\item $\hspace*{3mm}$ \textbf{for} $t=1,\ldots,T_{max}$ \textbf{do}
\item $\hspace*{6mm}$ \textbf{for} $j=1,\ldots,N_b$ \textbf{do}
\item $\hspace*{9mm}$ \textbf{for} each agent \textbf{do}
\item $\hspace*{12mm}$ Generate experiences tuple $(\mathbf{s}_j(t),\mathbf{a}_j(t),r_j(t),\mathbf{s}_j(t+1))$ based on Algorithm 1
\item $\hspace*{12mm}$ Store experiences to memory block $D_{\mathcal{T}_{j}}$
\item $\hspace*{12mm}$ Sample mini-batch of experiences $(\mathbf{s}_j,\mathbf{a}_j,\mathbf{r}_j,\mathbf{\tilde{s}}_j)$ from $D_{\mathcal{T}_{j}}$ as $\mathbb{D}_j^{tr}$
\item $\hspace*{12mm}$ \textbf{for} $i=1,\ldots,G_{in}$
\item $\hspace*{15mm}$ Update network parameter $\hat{\mathbf{\theta}}_j$, $\hat{\mathbf{\varphi}}_j$, and $\hat{\mathbf{\phi}}_j$ based on \eqref{innermore}
\item $\hspace*{12mm}$ \textbf{end for}
\item $\hspace*{9mm}$ Sample mini-batch experiences $(\mathbf{s}_j,\mathbf{a}_j,\mathbf{r}_j,\mathbf{\tilde{s}}_j)$ from $D_{\mathcal{T}_{j}}$ as $\mathbb{D}_j^{val}$
\item $\hspace*{9mm}$ Evaluate the gradient of the loss function of task on $\mathbb{D}_j^{val}$
\item $\hspace*{6mm}$ \textbf{end for}
\item $\hspace*{6mm}$ Update the global network parameter $\mathbf{\theta}$, $\mathbf{\varphi}$, and $\mathbf{\phi}$ by \eqref{globalupdate} or by using ADAM optimizer
\item $\hspace*{3mm}$ \textbf{end for}
\item \textbf{end for}
\end{enumerate}
\hrule
\vspace{2mm}
$\hspace*{4mm}$$\mathbf{Meta-adaptation}$
\vspace{-1mm}
\begin{enumerate}
\item Initialize $\hat{\mathbf{\theta}}_{ap}\leftarrow\mathbf{\theta}$, $\hat{\mathbf{\varphi}}_{ap}\leftarrow\mathbf{\varphi}$, and $\hat{\mathbf{\phi}}_{ap}\leftarrow\mathbf{\phi}$
\item \textbf{for} $t=1,\ldots,T_{Ap}$ \textbf{do}
\item $\hspace*{3mm}$ Generate experiences tuple $(\mathbf{s}(t),a^s(t),a^p(t),r(t),\mathbf{s}(t))$ based on Algorithm 1
\item $\hspace*{3mm}$ Store the experiences into the replay memory $\mathbb{D}_{ad}$
\item $\hspace*{3mm}$ Sample mini-batch of experiences and update parameters by \eqref{adaptmore} or using ADAM
\item \textbf{end}
\end{enumerate}
\hrule
\end{table}

\section{Simulation Results}\label{simu}
Numerical results are presented in this section to demonstrate the effectiveness of the proposed DRL and meta-based DRL algorithms in a single-cell V2X communications system.  Following the guideline about the simulation setup for V2X communications in 3GPP TR 36.885 \cite{3gpp2017technical}, we set the parameters for our simulator as: carrier frequency is 2 GHz; the number of V2I links is $M=4$; the antenna gain and receiver noise figure of the BS are 8 dBi and 5 dB, respectively; the antenna gain and receiver noise figure of vehicles are 3 dBi and 9 dB, respectively; the noise power is -114 dBm; the V2I transmit power is 35 dBm; the maximum tolerant latency for V2V links is 100 ms; and the vehicle antenna height is 1.5 m. The weights $v^i$, $v^v$, and $\lambda$ used in the immediate reward are 0.1, 0.9, and 1, respectively.

  Following the setup of DNN in V2X communications \cite{ye2019deep,liang2019spectrum}, a five-layer DNN is utilized to construct the neural network for DQL and DDPG algorithms, in which three hidden layers include 500, 250, and 120 neurons, respectively. The rectified linear unit (ReLU) is used as the activation function for the hidden layers used in both DQN and DDPG.  Sigmoid is used as the activation function for the output layer of the actor network of DDPG to scale the output into the transmission power region. The adaptive moment estimation method (Adam) is used to update the network parameters \cite{adam}. The payload $B$ for each V2V pair is 1060 bytes unless specified. At the beginning of each payload transmission, the remaining load equals to the payload for each V2V pairs.  Based on a trial-and-error approach, we consider to use 0.001 as the learning rate of the DQN unit and use 0.0001 and 0.001 as the learning rates of the actor and critic networks in DDPG, respectively. The numbers of samples in mini-batches used for the training of the proposed DRL algorithm and the proposed meta DRL algorithm are 2000 and 100, respectively. For the meta training, the number of tasks in each batch is 100, and the number of batches is 200. The learning rates used in the meta learning are set as $\hat{\alpha}=0.01$, $\hat{\beta}_c=0.01$, and $\hat{\beta}_a=0.001$. The discount factor is $\gamma=0.5$, and the update frequency factor (in (18) and (19)) of the target network in DDPG is $\tau=0.001$. All simulation results are generated by using PyTorch 1.4.0 on the Python 3.6 platform.

The following three communications scenarios are considered to demonstrate the efficiency of the proposed algorithms. The urban case is considered as the baseline communications scenario for all simulations in which the DRL model is trained.
\begin{itemize}
\item Urban case: the urban communications scenario introduced in 3GPP TR 22.886 \cite{3GPPTR37885} is considered in this case. The Manhattan layout is used to set up the environment, in which nine grids with a size of 250m $\times$ 433m for each grid are considered. Two lanes with 3.5 m street width in each direction are adopted. Both line-of-sight (LOS) and non-line-of-sight (NLOS) fading channels are considered. The log-normal distribution with 8 dB and 3 dB standard deviations is used to generate the shadowing for V2I and V2V links, respectively. The vehicle speed is set as 50 km/h.
\item Highway case: the freeway case introduced in 3GPP TR 22.886 \cite{3GPPTR37885} is used for the highway communications scenario setup. In this case, we consider 3 lanes with a width of 5 m in each direction and 120 km/h for all vehicles. The BS is located 35 m away from the freeway, which has the length of 1500 m. Only the LOS fading channel is considered in this case.
\item Rural case: the path loss and shadowing of the rural scenario in the WINNER II model is used \cite{bultitude20074}. The layout is similar to the urban case, but with wider street lanes (5 m) and a greater grid size (1000 m $\times$ 1000 m) as well as less occlusion from the buildings. The height of BS is set to 35 m. The speed of vehicles is 108 km/h.
\end{itemize}

In order to demonstrate the effectiveness of the proposed algorithms, we introduce three benchmarks as the comparison solutions, namely the DQN solution, the no-adaptation solution, and the random solution. The definitions of all solutions  are introduced as follows:
\begin{itemize}
\item The proposed DRL solution: this solution shows the result generated by the proposed DRL algorithm in Algorithm 1.  {In this solution, the communications scenario used for training and implementation are the same, which means that this solution shows the results without mismatch and thus provides a performance upper bound.}
\item The proposed meta-DRL solution: this solution shows the fast adaptation result of the proposed meta-RL algorithm in Algorithm 2.
\item The DQN solution: this solution shows the result for the algorithm that converts the continuous power to the discrete power with 5 quantization levels.  {The solution is trained and implemented based on the same communications scenario. It shows the results without mismatch.}
\item The no-adaptation solution: this solution shows the mismatch results for both DQN and the proposed DRL algorithms by which the model is trained in the baseline environment and tested in new environments (highway and rural cases).
\item The random solution: this solution shows the result that the sub-band and transmit power are randomly selected by each agent.
\end{itemize}
\subsection{Highway case}
\begin{figure}
\centering
\subfigure[]  
{
	\begin{minipage}{3.3in}
    \label{fig1:a}
    \centering
	\includegraphics[width=3.3in]{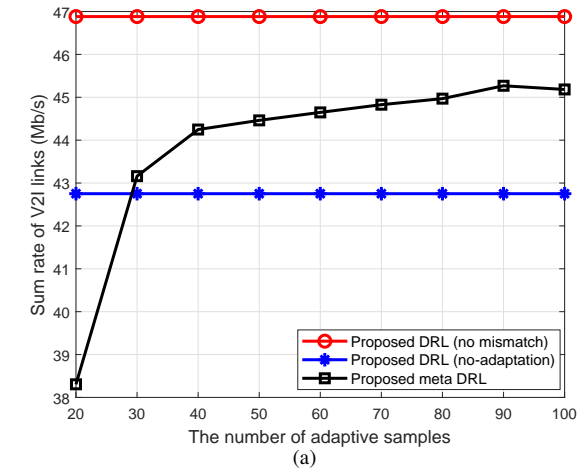}   

	\end{minipage}
}
\subfigure[]  
{
	\begin{minipage}{3.3in}
    \label{fig1:b}
    \centering
	\includegraphics[width=3.3in]{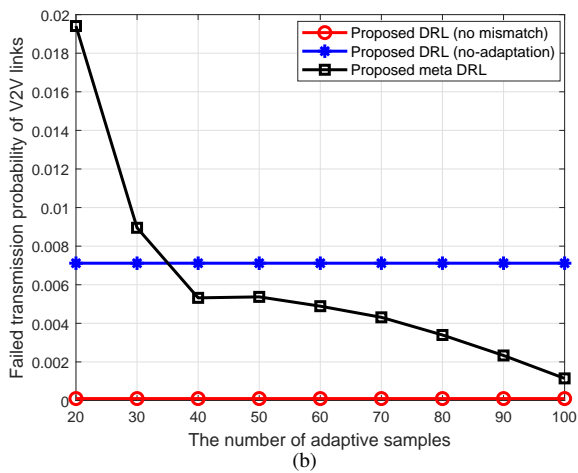}   
	\end{minipage}
}
\centering
\caption{The performance of adaptive samples for the proposed meta-RL algorithm for different metrics: (a) V2I sum rate (b) V2V transmission failure probability.}  
\label{fig1}   
\end{figure}
First, we will show the results comparison for the proposed algorithms in the highway case. \textcolor{black}{In this case, the model is trained based on the urban scenario and tested in the highway scenario.} All vehicles are uniformly distributed in the lanes for both directions. Obviously the agent performs better when it receives as much as possible experiences in a new environment. However, gathering a large amount of experiences takes a lot of time, which will be against the original intention of designing the meta-based DRL algorithm. In order to choose a proper adaptation sample numbers, an investigation is provided in Fig. \ref{fig1} to evaluate the effects of the number of received adaptation samples on the performance of the proposed meta-RL algorithm. For the simulation in Fig. \ref{fig1}, we define that each adaptive sample is generated at each time step and includes the experiences for all agent. As we can see from Fig. \ref{fig1}, the V2I sum rate increases and the failure probability of V2V links decreases as the number of adaptive samples increases. When the number of adaptive samples is greater than 30 and 40, the V2I sum rate and the V2V failure probability generated by the proposed meta-RL algorithm outperform the no-adaptation solution, respectively. In order to consider the tradeoff between the overhead and performance, we use 40 samples to perform the adaptation in the remaining of the simulations.

\begin{figure}
\centering
\subfigure[]  
{
	\begin{minipage}{3.3in}
    \label{fig2:a}
    \centering
	\includegraphics[width=3.3in]{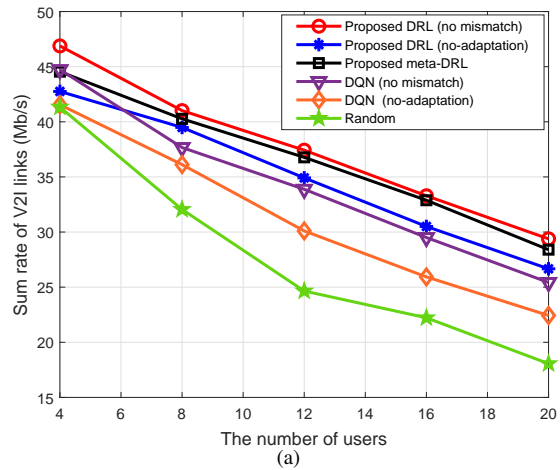}   

	\end{minipage}
}
\subfigure[]  
{
	\begin{minipage}{3.3in}
    \label{fig2:b}
    \centering
	\includegraphics[width=3.3in]{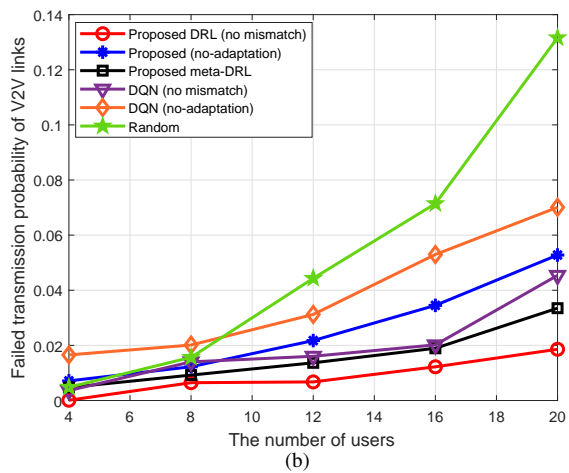}   
	\end{minipage}
}
\centering
\caption{The performance comparison for different solutions in the highway case via two metrics: (a) V2I sum rate (b) V2V transmission failure probability.}  
\label{fig2}   
\end{figure}
After investigating the effects of sample numbers, we evaluate the proposed DRL and meta-based DRL algorithms in the highway case. The effectiveness of the proposed algorithms is demonstrated by comparing with benchmark solutions under two different metrics: the sum rate of V2I links in Fig. \ref{fig2:a} and the failure transmission probability of V2V links in Fig. \ref{fig2:b}. As can be seen, the sum rate of V2I links decreases and the failure transmission probability of V2V links increases as the number of users for all solutions. \textcolor{black}{This is because when the number of users increases, more V2V links will share the fixed number of sub-bands, which causes   higher interference to the V2I links and V2V links.} From Fig. \ref{fig2:a}, the proposed DRL algorithm achieves a better gain on the sum rate of V2I links compared to the DQN algorithm, which uses the quantization method to quantize the continuous transmit power. The reason is that quantization will cause the performance loss. Both DRL algorithms outperform  the random solution. There exists an obviously gap between the normal training without mismatch and the non-adaptation training. The reason is because the model used in the non-adaptation solution is trained in the  urban scenario. This indicates that although the proposed DRL algorithm can provide efficient decisions on the resource allocation, it cannot overcome the negative effects caused by the environment change. Hence, it is important to design an algorithm which is able to eliminate the negative effects of mismatch thus avoiding training the policy from scratch. \textcolor{black}{The proposed meta-based DRL algorithm generates the sum rate very close to the proposed DRL algorithm (without mismatch), and more importantly it only needs 40 samples to achieve such performance compared to more than 3000 samples in training from scratch.} The proposed algorithms provide better performance not only for the V2I links but also for the V2V links. The proposed DRL algorithm outperforms the DQN and random solutions. In addition, the proposed meta-DRL algorithm can significantly reduce the failure probability of V2V links compared to the non-adaptation solution.
\subsection{Rural case}
\begin{figure}
\centering
\subfigure[]  
{
	\begin{minipage}{3.3in}
    \label{fig3:a}
    \centering
	\includegraphics[width=3.3in]{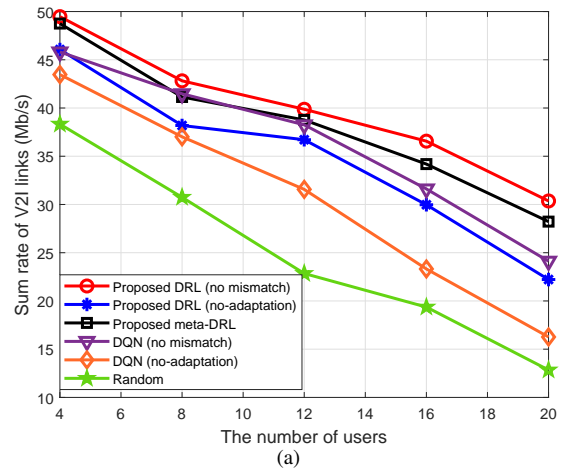}    
	\end{minipage}
}
\subfigure[] 
{
	\begin{minipage}{3.3in}
    \label{fig3:b}
    \centering
	\includegraphics[width=3.3in]{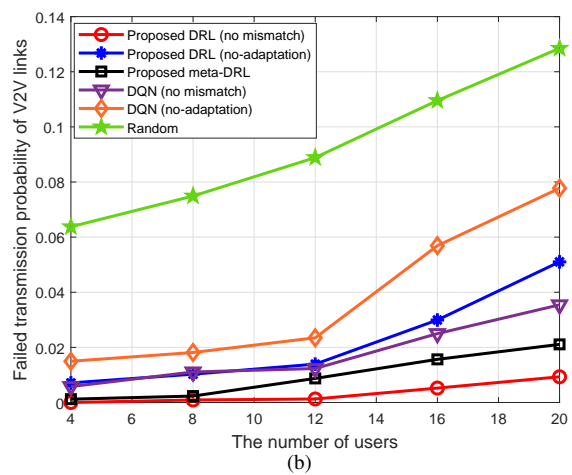}    
	\end{minipage}
}
\centering
\caption{The performance comparison for different solutions in the rural case via two metrics: (a) V2I sum rate (b) V2V transmission failure probability.}  
\label{fig3}   
\end{figure}
After evaluating the performance of algorithms in the highway scenario, we investigate the effectiveness of the proposed algorithms in the rural vehicular communications scenario. In this case, the model is still trained in the urban scenario. Fig. \ref{fig3} shows the performance of V2I links and V2V links for different solutions. As can be seen from Fig. \ref{fig3:a}, the proposed RL algorithm still achieves the best sum rate performance for V2I links compared to the benchmark solutions. The random solution performs worst since randomly selecting the sub-band and power will cause  more interference. The proposed meta-RL algorithm can provide efficient and fast adaptation when the communications environment changes. For the results of V2V links in Fig. \ref{fig3:b}, the proposed DRL and meta-based DRL algorithm can reduce the failure probability, especially when   the  number of vehicles is large. It indicates that the proposed algorithms have good interference management ability.
 
\subsection{Urban case}
\begin{figure}
\centering
\subfigure[]  
{
	\begin{minipage}{3.3in}
    \label{fig4:a}
    \centering
	\includegraphics[width=3.3in]{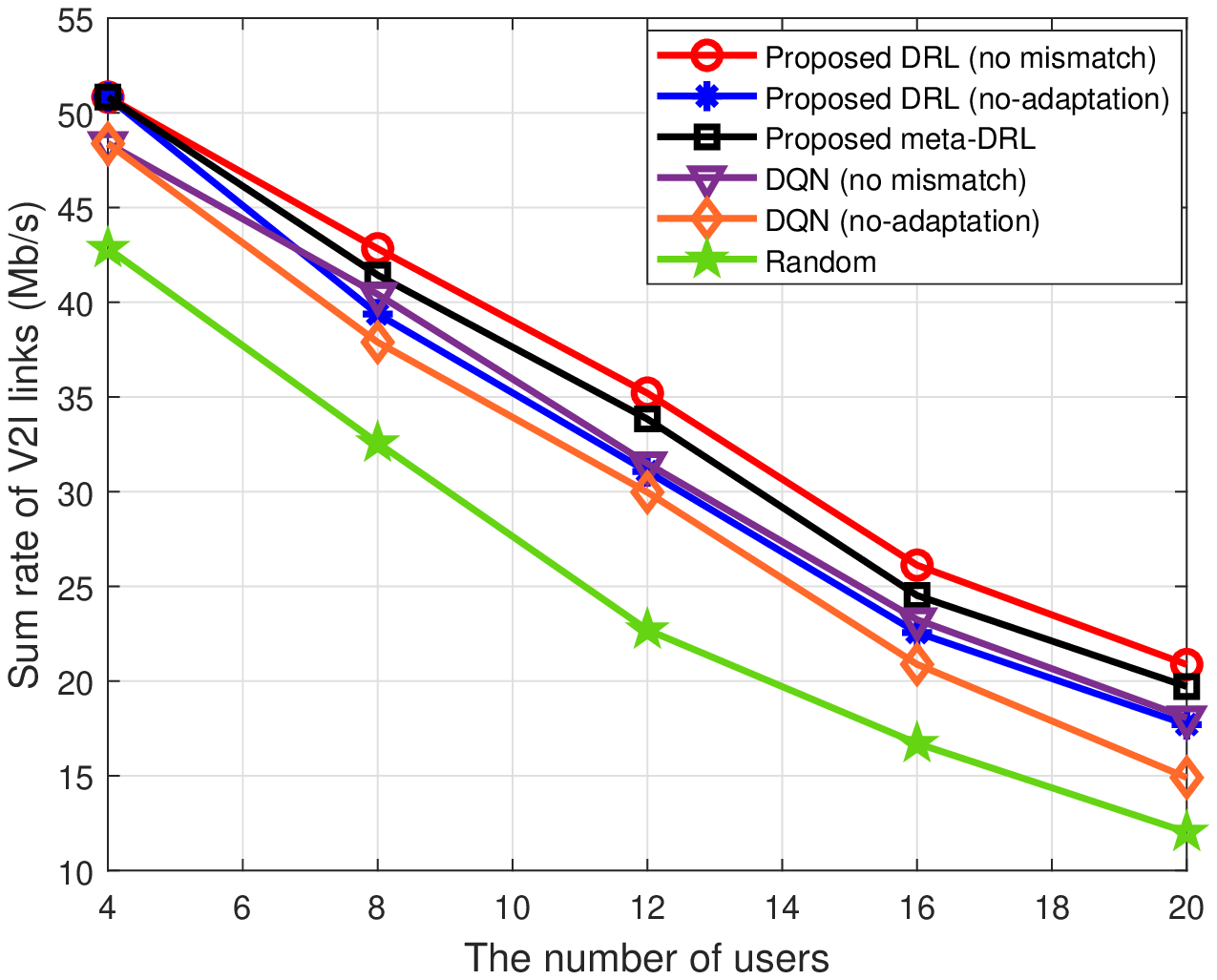}    

	\end{minipage}
}
\subfigure[]  
{
	\begin{minipage}{3.3in}
    \label{fig4:b}
    \centering
	\includegraphics[width=3.3in]{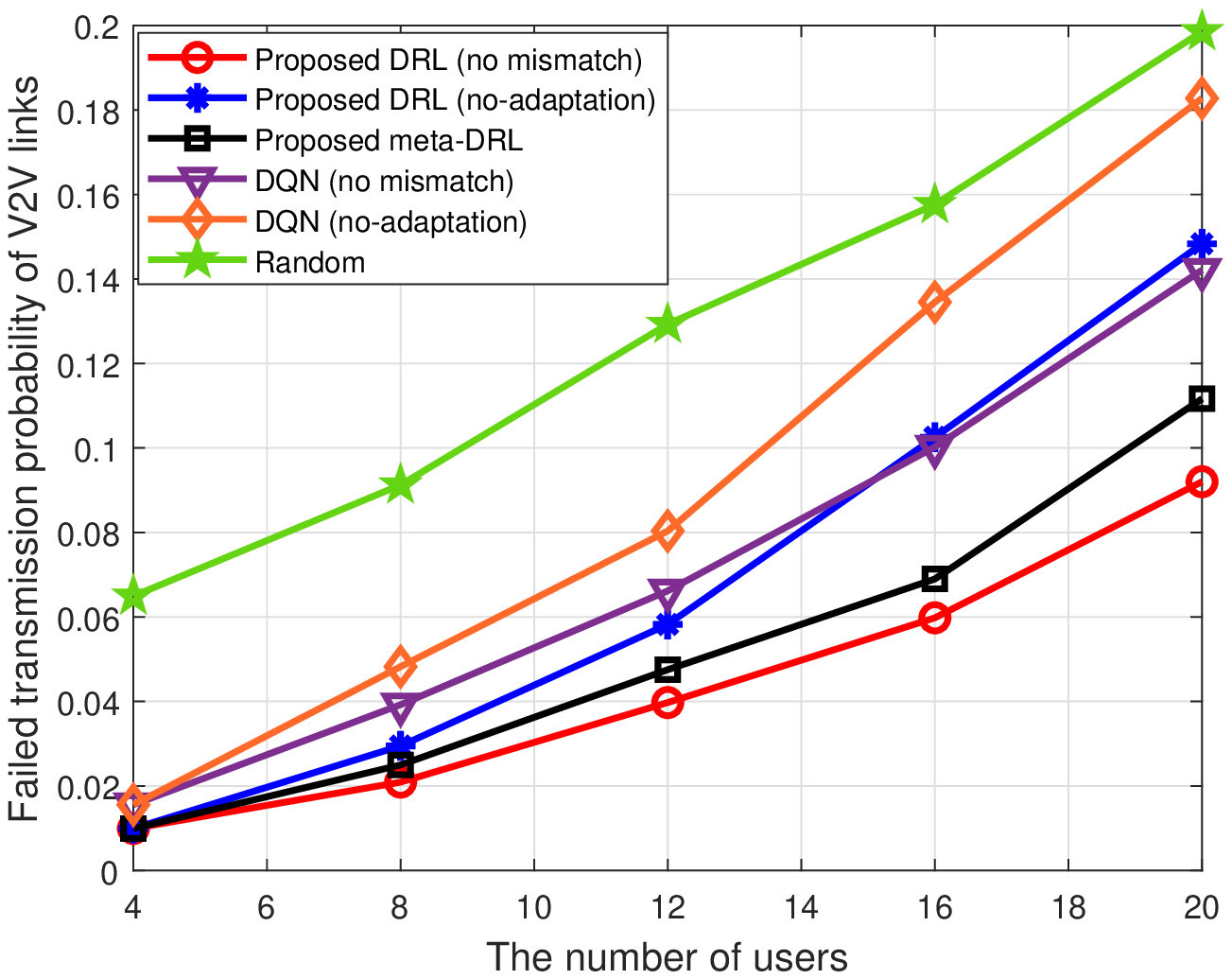}    
	\end{minipage}
}
\centering
\caption{The performance comparison for different solutions versus the number of vehicles in the urban case via two metrics: (a) V2I sum rate (b) V2V transmission failure probability.}  
\label{fig4}   
\end{figure}

\begin{figure}
\centering
\subfigure[]  
{
	\begin{minipage}{3.3in}
    \label{fig5:a}
    \centering
	\includegraphics[width=3.3in]{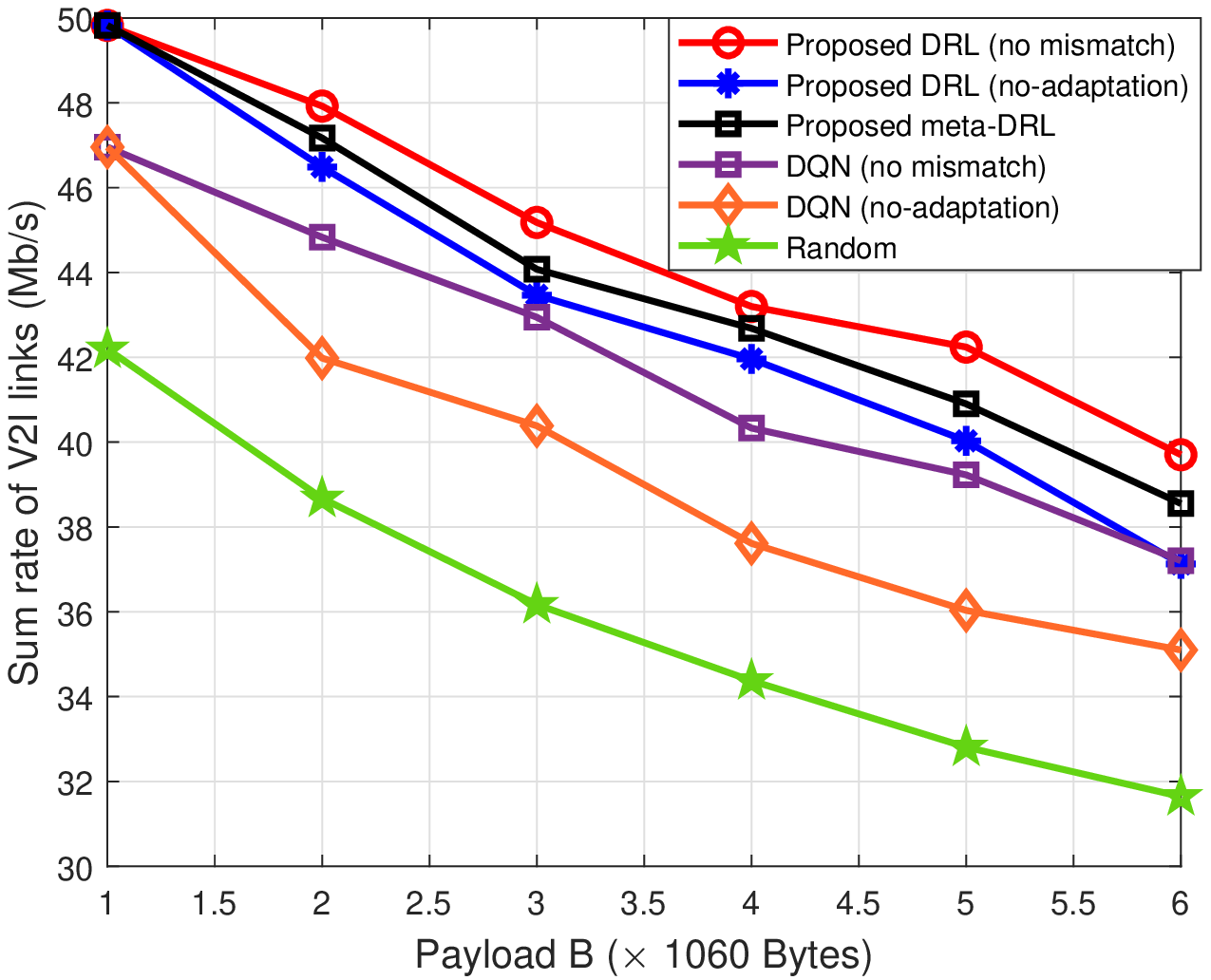}    

	\end{minipage}
}
\subfigure[]  
{
	\begin{minipage}{3.3in}
    \label{fig5:b}
    \centering
	\includegraphics[width=3.3in]{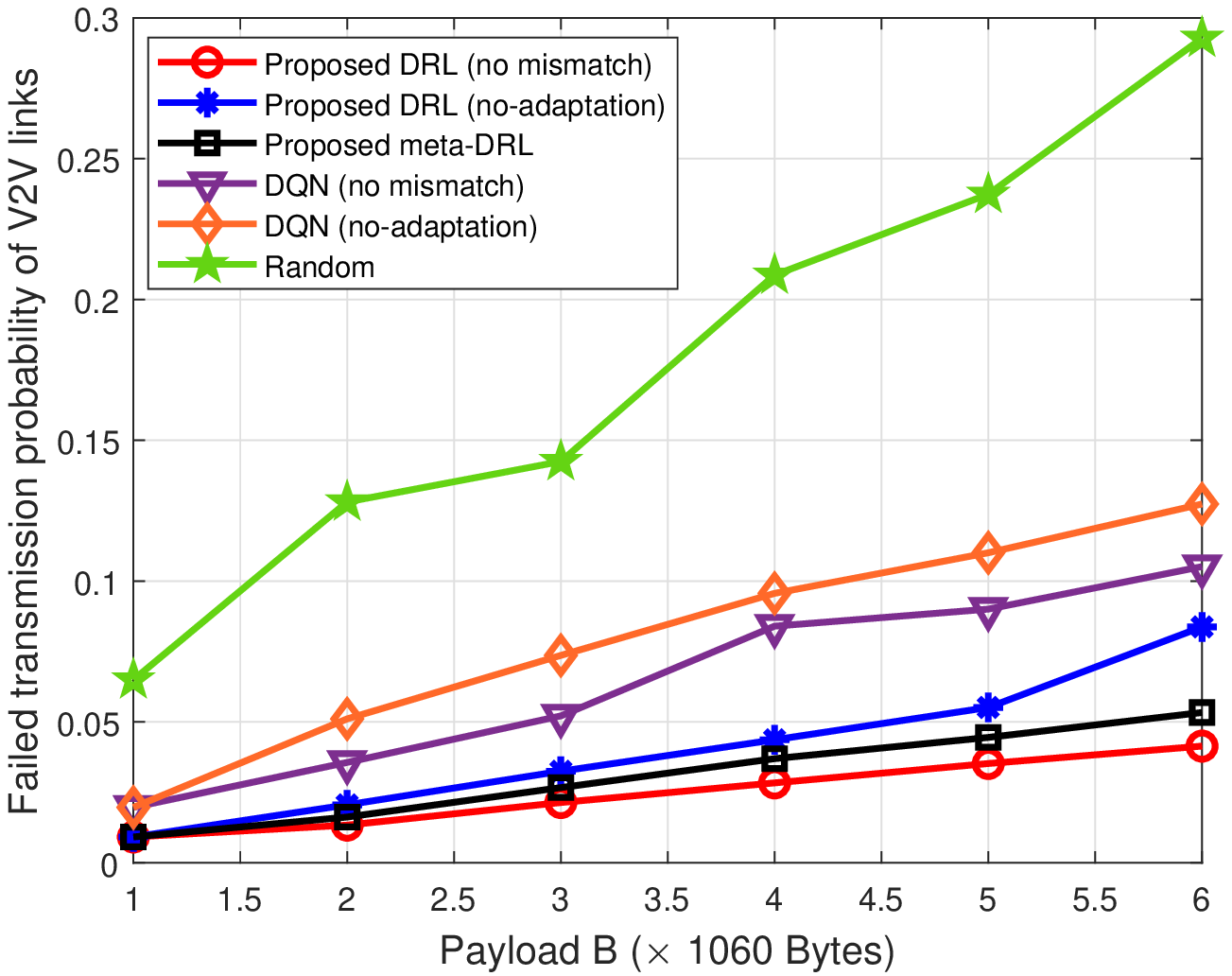}    
	\end{minipage}
}
\centering
\caption{The performance comparison for different solutions versus V2V payload in the urban case via two metrics: (a) V2I sum rate (b) V2V transmission failure probability.}  
\label{fig5}   
\end{figure}

The results plotted in Fig. \ref{fig2} and Fig. \ref{fig3} have demonstrated the effectiveness of the proposed algorithms on resource allocation and generalization ability in different V2X scenarios. In order to gain more insight on whether the proposed algorithms can still provide similar performance gain in the same environment when other parameters change, we examine the proposed algorithms in the urban case with varying number of vehicles in Fig. \ref{fig4} and V2V payload in Fig. \ref{fig5}. \textcolor{black}{The baseline model is trained by using the case with four vehicles in Fig. \ref{fig4} and by using the payload with 1060 bytes in Fig. \ref{fig5}.}
Fig. \ref{fig4} shows the V2I and V2V performance with respect to the increasing  number of vehicles for different resource allocation solutions. According to Fig. \ref{fig4:a}, the V2I sum rate drops for all solutions with increasing number of vehicles. This is due to the increased V2V interference to V2I links. The proposed DRL algorithm still achieves the highest sum rate compared to the DQN and random solutions. The mismatch gap appears between the adaptation and non-adaptation solution when the number of vehicle used for testing increases. The proposed meta-based DRL algorithm achieves the expected results, which are more closer to the proposed DRL solution compared to the non-adaptation solution. The expected results are also obtained in the V2V performance in Fig. \ref{fig4:b}.

Next, we look into the effects of different V2V payload sizes on the performance of the proposed algorithms. Fig. \ref{fig5} presents the performance of the sum rate and the transmission failure probability versus the different V2V payload sizes for different solutions. From the figure we can see that the V2I sum rate drops and the V2V transmission failure probability increases with growing V2V payload sizes. This is because the increasing payload requires longer transmission duration and higher transmit power for each V2V link, which will cause stronger interference to the V2I and V2V links. Compared to the DQN solution, the proposed DRL algorithm is able to achieve higher V2I sum rate with lower V2V failure probability when the V2V payload increases. This observation demonstrates the robustness of the proposed DRL algorithm   regarding the V2V payload. In addition, the proposed meta-based DRL algorithm achieves the expected results in terms of the V2I sum rate and the V2V transmission failure probability. This fact verifies the effectiveness and robustness of the proposed meta-based DRL algorithm on solving the resource allocation problem with mismatch issues.
\section{Conclusions}\label{conc}
This paper considered efficient decision-making policy for resource allocations for V2X communications in both stationary and dynamic environments. To achieve this goal, we formulated a sub-band assignment and power allocation problem as an MDP, which aims to maximize the sum rate of V2I links while satisfying the latency requirement of V2V links. A decentralized joint DRL-based algorithm was proposed to solve the problem by using DQN for the sub-band assignment and using DDPG for the transmit power allocation. In order to increase the adaptation ability of the proposed DRL algorithm in the dynamic environment, we further proposed a meta-based DRL algorithm by combining meta-learning and DRL. Compared to the DQN-based algorithm, our DRL-based algorithm can provide better performance for both V2I and V2V links. In addition, the policy trained by using the proposed meta-based DRL algorithm has a good generalization ability and can fast adapt to new environments via limited experiences. As to future work, it would be an interesting direction to extend our decision-making strategy to V2X communication scenarios with multiple antennas in which the effect of beamforming on the resource allocation will be considered. In addition, to design the efficient decision-making algorithms for the vehicular networks with imperfect CSI is another future direction. Furthermore, our proposed algorithms can be used as a framework to design specific algorithms for other communications scenarios with   mismatch issues, such as mobile wireless sensor networks and D2D networks.

 \end{document}